\documentclass[onecolumn, 12pt]{IEEEtran}
\usepackage{amsmath,amsfonts,amssymb,enumerate,cite}
\usepackage[final]{lscape,graphicx,epsfig}
\usepackage{cite}
\DeclareMathOperator*{\argmax}{argmax}
\DeclareMathOperator*{\argmin}{argmin}

\DeclareMathOperator*{\maximize}{maximize}

\newtheorem{thm}{Theorem}

\newtheorem{corollary}{Corollary}

\title{Distributed Detection over Fading MACs with Multiple Antennas
at the Fusion Center}

\author{Mahesh K. Banavar, {\em Student Member,
IEEE}, Anthony D. Smith, Cihan Tepedelenlio\u{g}lu, {\em Member,
IEEE}, and Andreas Spanias, {\em Fellow, IEEE} \thanks{The work in
this paper is supported by the SenSIP Center, Arizona State
University. A.D. Smith is with L-3 Communications. M.K. Banavar, C.
Tepedelenlio\u{g}lu and A. Spanias are with the SenSIP Center,
School of Electrical, Computer and Energy Engineering, Fulton
Schools of Engineering, Arizona State University, Tempe, AZ 85287.
USA. (Email: maheshkb@asu.edu, tony.d.smith@l-3com.com, \{cihan,
spanias\}@asu.edu). }}

\date{}

\begin{document}

\maketitle
\bibliographystyle{IEEEtran}

\begin{abstract}
A distributed detection problem over fading Gaussian multiple-access
channels is considered. Sensors observe a phenomenon and transmit
their observations to a fusion center using the amplify and forward
scheme. The fusion center has multiple antennas with different
channel models considered between the sensors and the fusion center,
and different cases of channel state information are assumed at the
sensors. The performance is evaluated in terms of the error exponent
for each of these cases, where the effect of multiple antennas at
the fusion center is studied. It is shown that for zero-mean channels
between the sensors and the fusion center when there is no
channel information at the sensors, arbitrarily large gains in the
error exponent can be obtained with sufficient increase in the
number of antennas at the fusion center. In stark contrast, when
there is channel information at the sensors, the gain in error
exponent due to having multiple antennas at the fusion center is
shown to be no more than a factor of $8/\pi$ for Rayleigh fading
channels between the sensors and the fusion center, independent
of the number of antennas at the fusion center, or correlation
among noise samples across sensors. Scaling laws for such gains are
also provided when both sensors and antennas are increased
simultaneously. Simple practical schemes and a numerical method using semidefinite relaxation techniques are presented that utilize
the limited possible gains available. Simulations are used to
establish the accuracy of the results.
\end{abstract}

\section{Introduction}
\label{sec:introduction}

Sensors are becoming commonplace in factories, environmental, and
home appliance monitoring, as well as in scientific study. In many
such applications, a number of independent sensors each make a local
observation, which are transmitted to a fusion center (FC) after
limited initial processing at the sensors, and combined at the FC to
calculate a global decision \cite{viswanathan}. Sensors may adopt
either a digital or an analog method for relaying the sensed
information to the FC. The digital method consists of quantizing the
sensed data and transmitting the digital data over a
rate-constrained channel \cite{xiao05}. In these cases, the required
channel bandwidth is quantified by the number of bits being
transmitted between the sensors and the FC. In contrast, the analog
method consists of amplifying and then forwarding the sensed data to
the FC, while respecting a power constraint \cite{banavar09, Chamberland2004}. The
transmissions can be appropriately pulse-shaped and amplitude
modulated to consume finite bandwidth. The channels between the
sensors and the FC can be orthogonal, in which case, the
transmissions from each sensor are separately received at the FC
\cite{xiao05}. On the other hand, with multiple-access channels
between the sensors and the FC, the noisy sum of all the
transmissions are received at the FC to make a decision \cite{Chen04, Li07, Wimalajeewa08,
Evans08a, banavar09}. The bandwidth
requirements of sensor networks with orthogonal channels scale
linearly with the number of sensors, whereas, when the channels are
multiple-access, transmissions are simultaneous and in the same
frequency band, keeping the utilized bandwidth independent of the
number of sensors in the network. 

Distributed detection problems have been mainly studied assuming a
single receive antenna at the FC. It is possible that introducing
multiple antennas at a receiver may overcome the degradations caused
by multi-path fading and noise. Inspired by conventional MIMO
systems, a natural question is how much performance gain can be
expected from adding multiple antennas at the FC in a distributed
detection problem. However, this question cannot be directly
answered by the studies in the MIMO literature. Adding multiple
antennas to the FC for distributed detection problems is different
when compared to the analysis of conventional MIMO systems for two
reasons: (i) the presence of sensing noise (the parameter of
interest is corrupted {\em before} transmission); and (ii) a large
number of sensors enable {\em asymptotic} analysis. To the best of
our knowledge, references \cite{Chen96, Zhang08} are the only works
that consider multiple antennas at the FC. In \cite{Chen96}, a
decision fusion problem with binary symmetric channels between the
users and the FC is considered where the data are quantized at the
sensors, transmitted over parallel channels, and processed after
being received by three antennas. In \cite{Zhang08}, the authors
consider multiple antennas at the FC. However, they consider a set
of deterministic gains for the orthogonal channels, known at the
sensors. They do not consider multiple-access channels, or
characterize the performance benefits of adding antennas at the FC
in the presence of fading. The system models in \cite{Liu2007,
Anandkumar2007, Berger2009, Yiu2009, Jayaweera05, Jayaweera07} are similar to adding multiple
antennas at the FC, where the authors consider other forms of
diversity, such as independent frequencies, CDMA codewords or several time intervals
over fast-time-varying channels. The main difference between these
papers and our results is the fact that we use asymptotic techniques
to investigate the benefits of adding multiple antennas at the
fusion center, when the number of sensors grows large. We show that
the gain on the error exponent by adding antennas to the FC when
there is no CSI at the sensors grows linearly with the number of
antennas. In stark contrast, when there is CSI at the sensors, only
limited gains are possible by adding antennas at the FC. This is
unlike what we see in traditional MIMO wireless communications,
where adding antennas at the FC will result either in diversity gain
or array gain, for asymptotically large SNRs.

\begin{figure}[tb]
\begin{minipage}[b]{1.0\linewidth}
  \centering
  \centerline{\epsfig{figure=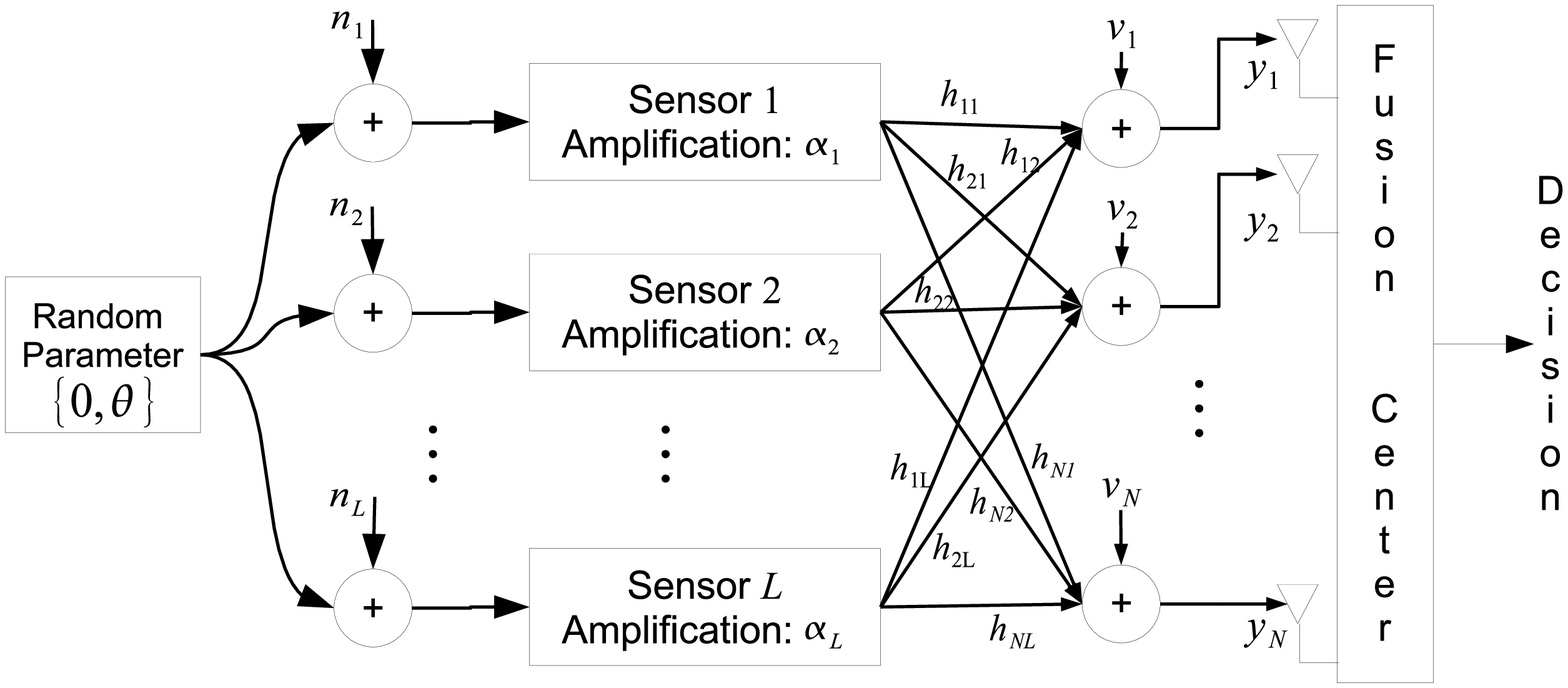,width=8cm,height=4cm}}
\end{minipage}
\caption{System Model: A random parameter is sensed by $L$ sensors. Each sensor transmits amplified observations over fading multiple access channels to a fusion center with $N$ antennas.} \label{Fig:problem_setup}
\end{figure}

In this paper, a distributed detection problem over a multiple
access channel, where the FC has multiple antennas is considered
(Figure \ref{Fig:problem_setup}). The data collected by the sensors
are transmitted to the FC using the amplify and forward scheme, with
a total power constraint on the sensor gains. Performance is
evaluated when the sensors have no channel information, have full
channel information and partial channel information in the presence
of fading, both with zero and non-zero mean. Analysis is performed
for two cases: (a) large number of sensors and a fixed number of
antennas, and (b) large number of antennas {\em and} sensors with a
fixed ratio. In each case, the error exponent is used as the metric
to quantify performance through the effect of channel statistics and
the number of antennas. It is shown that the system performance
depends on the channel distribution through its first and second
order moments. This information is used to address {\em our main
objective}, which is to quantify the gain possible by adding
multiple antennas at the FC over fading multiple-access channels for
distributed detection problems.

\section{System Model}
\label{sec:model}

A sensor network, illustrated in Figure \ref{Fig:problem_setup},
consisting of $L$ sensors and a fusion center with $N$ antennas is
considered. The sensors are used to observe a parameter $\Theta \in
\{ 0, \theta \}$. The value, $x_{l}$, observed at the $l^{th}$
sensor is
\begin{equation}
x_{l} =
\begin{cases}
\eta_{l} & \text{under $H_{0}$}\\
\theta + \eta_l & \text{under $H_{1}$}
\end{cases}
\label{eqn:obs_eqn}
\end{equation}
for $l = 1,...,L$. It is assumed that $\eta_{l} \sim \mathcal{CN}
(0, \sigma_{\eta}^{2})$ are iid, the hypothesis $H_{1}$ occurs with
{\em a priori} probability, $0 < p_{1} < 1$, and the hypothesis
$H_{0}$ with probability $p_{0} = 1-p_{1}$. The $l^{th}$ sensor
applies a complex gain, $\alpha_l$, to the observed value, $x_l$.
This amplified signal is transmitted from sensor $l$ to antenna $n$
over a fading channel, $h_{nl}$, $n = 1, ..., N$, and $l = 1, ...,
L$, which are iid and satisfy $E[|h_{nl}|^{2}] = 1$. Unless
otherwise specified, no other assumptions are made on the channel
distribution. The $n^{th}$ antenna receives a superposition of all
sensor transmissions in the presence of iid channel noise,
$\nu_n\sim\mathcal{CN}(0,\sigma_{\nu}^{2})$, such that
\begin{equation}
y_{n} = \sum_{i=1}^{L} h_{in}\alpha_i(\Theta + \eta_i) + \nu_n,
\label{eqn:yn}
\end{equation}
where $\{\eta_{i}\}_{i=1}^{L}$ and $\{\nu_{n}\}_{n=1}^{N}$ are
independent.

Defining {\boldmath$\alpha$} as an $L\times 1$ vector containing
$\{\alpha_{i}\}_{i=1}^{L}$, $\mathbf{D}(\boldsymbol{\alpha})$ an
$L\times L$ diagonal matrix with the components of
{\boldmath$\alpha$} along the diagonal, the received signal is
expressed in vector form as
\begin{equation}
\mathbf{y} = \mathbf{H}\boldsymbol{\alpha}\Theta +
\mathbf{HD}(\boldsymbol{\alpha}) \boldsymbol{\eta} +
\boldsymbol{\nu}, \label{eqn:y_vec_init}
\end{equation}
where $\mathbf{H}$ is an $N \times L$ matrix containing the elements
$h_{nl}$ in the $n^{th}$ row and $l^{th}$ column,
$\boldsymbol{\eta}$ is an $L\times 1$ vector containing
$\{\eta_{i}\}_{i=1}^{L}$, and $\boldsymbol{\nu}$ is an $N\times 1$
vector containing $\{\nu_{n}\}_{n=1}^{N}$. Based on the received signal, $\mathbf{y}$ (from
(\ref{eqn:y_vec_init})), the FC decides on one of the two hypotheses
$H_{0}$ or $H_{1}$. Since the FC has full knowledge of $\mathbf{H}$
and $\boldsymbol{\alpha}$, $\mathbf{y}$ is Gaussian distributed
under both hypotheses:
\begin{align}
H_{0}: \mathbf{y} & \sim \mathcal{CN}(\mathbf{0}_{N},\mathbf{R}
(\boldsymbol{\alpha}))
\nonumber \\
H_{1}: \mathbf{y} & \sim \mathcal{CN}(\theta
\mathbf{H}\boldsymbol{\alpha},\mathbf{R} (\boldsymbol{\alpha}))
\label{eqn:hyp_gauss}
\end{align}
where $\mathbf{0}_{N}$ is an $N\times1$ vector of zeros and
$\mathbf{R}(\boldsymbol{\alpha})$ is the $N\times N$ covariance
matrix of the received signal given by
\begin{equation}
\mathbf{R} (\boldsymbol{\alpha}) = \sigma_{\eta}^{2} \text{ }
\mathbf{HD} (\boldsymbol{\alpha}) \mathbf{D}
(\boldsymbol{\alpha})^{H} \mathbf{H}^{H} + \sigma_{\nu}^{2} \text{ }
\mathbf{I}_{N}. \label{eqn:cov_R}
\end{equation}
We consider detection at a single snapshot in time, and therefore,
we do not have a time index.

\subsection{Power Constraint}
\label{ssec:power_constraint}

The $i^{th}$ sensor transmits $\alpha_{i}(\Theta + \eta_{i})$. The
total transmitted power is given by
\begin{equation}
P_{T} = E \left[ \sum_{i=1}^{L} \left| \alpha_{i}(\Theta + \eta_{i})
\right|^{2} \right] = \left( p_{1}\theta^{2} + \sigma_{\eta}^{2}
\right) \sum_{i=1}^{L} \left|\alpha_{i}\right|^{2}.
\label{eqn:power_constraint_setup}
\end{equation}
It should also be noted here that the instantaneous transmit power
from the sensors is $|\alpha_{i} (\theta + \eta_{i})|^{2}$. This is
a function of the actual realizations of sensing noise, making it
difficult to predict and constrain. Therefore, we constrain
$\alpha_{i}$'s, which allows imposing an average (over sensing
noise) power constraint. The sensor gains,
$\left\{\alpha_{i}\right\}$, are constrained by
\begin{equation}
P \mathop{:=} \sum_{i=1}^{L} |\alpha_{i}|^{2} =
\frac{P_{T}}{p_{1}\theta^{2} + \sigma_{\eta}^{2}}.
\label{eqn:power_constraint}
\end{equation}

\subsection{The Detection Algorithm and its Performance}
\label{ssec:err_exp}

Given the received data, $\mathbf{y}$, the FC selects the
appropriate hypothesis according to
\begin{equation}
\Re\{\theta \mathbf{y}^{H}\mathbf{R} (\boldsymbol{\alpha})^{-1}
\mathbf{H} \boldsymbol{\alpha}\}
\overset{H_{1}}{\underset{H_{0}}{\gtrless}} \frac{1}{2} \theta^{2}
\boldsymbol{\alpha}^{H} \mathbf{H}^{H} \mathbf{R}
(\boldsymbol{\alpha})^{-1} \mathbf{H} \boldsymbol{\alpha} + \tau,
\label{eqn:choose_H1}
\end{equation}
where $\tau$ is a threshold that can be selected using the
Neyman-Pearson or the Bayesian approach. Using (\ref{eqn:hyp_gauss})
and (\ref{eqn:choose_H1}), and the Bayesian test with the detection
threshold, $\tau = (1/2) \ln(p_{0}/p_{1})$, the probability of error
conditioned on the channel can be calculated as
\begin{equation}
P_{e|\mathbf{H}}(N) = p_{0} Q\left( \omega +
\tau/\omega \right) + p_{1} Q\left( \omega - \tau/\omega \right), \label{eqn:Pe}
\end{equation}
where $\omega \mathop{:=} \theta \sqrt{\boldsymbol{\alpha}^{H} \mathbf{H}^{H} \mathbf{R}
(\boldsymbol{\alpha})^{-1} \mathbf{H} \boldsymbol{\alpha}/2}$ for brevity, $N$ is the number of antennas at the FC and $Q(x) = \int_{x}^{\infty} \frac{1}{\sqrt{2\pi}} e^{-y^{2}/2}
dy$.  The error exponent is defined in terms of the conditional
error probability for the FC with $N$ antennas as \cite{Jayaweera05, Tarzai05, Bai08} 
\begin{equation}
\mathcal{E}(N) = \lim_{L\to\infty} -\frac{1}{L} \log
P_{e|\mathbf{H}}(N). \label{eqn:err_exp_defn}
\end{equation}
Note that even though $P_{e|\mathbf{H}}(N)$ in (\ref{eqn:Pe}) is a
channel-dependent random variable, we will show that the limit in
(\ref{eqn:err_exp_defn}) converges in probability to a deterministic
constant for the cases of interest to us. Substituting
(\ref{eqn:Pe}) into (\ref{eqn:err_exp_defn}), using L'H\^{o}pital's
rule, and the Leibniz Integral rule for differentiating under the
integral sign,
\begin{equation} \mathcal{E}(N) = \lim_{L\to\infty} \frac{1}{8}
\frac{1}{L}
\theta^{2}\boldsymbol{\alpha}^{H}\mathbf{H}^{H}\mathbf{R}
(\boldsymbol{\alpha})^{-1} \mathbf{H}\boldsymbol{\alpha}
\label{eqn:err_exp_fn}
\end{equation}
in probability, which does not depend on $p_{0}$ and $p_{1}$. Since
$\mathcal{E}(N)$ is the negative exponent of the probability of
error, a larger value represents better performance. The error
exponent in (\ref{eqn:err_exp_fn}) is a deterministic performance
metric over fading channels and depends on fading statistics. It can
also be viewed as a ``generalized SNR'' expression in this system
with multiple sensor and channel noise sources. We follow \cite{Jayaweera05, Tarzai05, Bai08} in our definition of the error exponent in (\ref{eqn:err_exp_defn}). Alternatively, one can consider the unconditional error exponent, $E_{\mathbf{H}} [P_{e|\mathbf{H}} (N)]$, which would depend on the distribution of $\mathbf{H}$ in (\ref{eqn:err_exp_defn}), in place of $P_{e|\mathbf{H}} (N)$. We will not pursue this approach herein. 

Our primary focus
throughout this paper is the dependence of (\ref{eqn:err_exp_fn}) on
\begin{enumerate}[(i)]
 \item the number of antennas, $N$, for different fading-channel
 distributions;
 \item different assumptions about the dependence of the sensor
 gains, $\boldsymbol{\alpha}$, on the channel, $\mathbf{H}$.
\end{enumerate}

With the Neyman-Pearson test, rather than the Bayesian test, it can
be shown that the error exponent is given by $\lim_{L\to\infty} 0.5
L^{-1} \theta^{2} \boldsymbol{\alpha}^{H} \mathbf{H}^{H} \mathbf{R}
(\boldsymbol{\alpha})^{-1} \mathbf{H} \boldsymbol{\alpha}$, which
does not depend on the false alarm probability and is a factor of
four greater than the error exponent derived in the Bayesian case.
Since the two cases differ only by a fixed constant, the Bayesian
approach will be used throughout.

\section{Performance over AWGN channels}
\label{sec:AWGN}

The error exponent with AWGN channels is computed to establish a
benchmark for the fading case of the next section, which is our main
focus. For AWGN channels, $h_{nl} = 1$. Due to symmetry and to
respect the power constraint, $\alpha_{i} = \sqrt{P/L}, \forall i$.
Defining $\mathbf{1}_{L}$ as an $L\times1$ vector of ones, and
$\mathbf{1}_{N\times L}$ as an $N\times L$ matrix of ones, we have
$\boldsymbol{\alpha} = \sqrt{P/L} \text{ } \mathbf{1}_{L}$ and
$\mathbf{H} = \mathbf{1}_{N\times L}$. Substituting these in
(\ref{eqn:cov_R}),
\begin{equation}
\mathbf{R} \mathbf{:=} \mathbf{R} (\sqrt{P/L} \mathbf{1}_{L}) =
\sigma_{\eta}^{2} P \mathbf{1}_{N\times N} + \sigma_{\nu}^{2}
\mathbf{I}_{N}. \label{eqn:R_AWGN}
\end{equation}
The inverse of (\ref{eqn:R_AWGN}) can be expressed using the
Sherman-Morrison-Woodbury formula for matrix inversion and
substituted into (\ref{eqn:err_exp_fn}) to yield
\begin{equation}
\mathcal{E}_{\rm AWGN}(N) \mathop{:=} \frac{1}{8} \frac{N
\gamma_{\rm s} \gamma_{\rm c} } {N \gamma_{\rm c} + p_{1}
\gamma_{\rm s} + 1}, \label{eqn:E_AWGN}
\end{equation}
where the sensing SNR is defined as $\gamma_{\rm s} \mathop{:=}
\theta^{2}/ \sigma_{\eta}^{2}$, and the channel SNR, $\gamma_{\rm c}
\mathop{:=} P_{T}/\sigma_{\nu}^{2}$. Since the partial derivative $\partial
\mathcal{E}_{\rm AWGN}(N)/\partial N >0$, for the AWGN case, having
multiple antennas improves the error exponent which can be
interpreted as array gain on the channel SNR $\gamma_{\rm c}$. As a
special case, consider $N=1$, to get the result for the single
antenna case:
\begin{equation}
\mathcal{E}_{\rm AWGN}(1) = \frac{1}{8} \frac{\gamma_{\rm c}
\gamma_{\rm s}} {\gamma_{\rm c} + p_{1} \gamma_{\rm s} + 1}.
\label{eqn:E_AWGN_1}
\end{equation}
With $p_{1} = 0.5$, $\gamma_{\rm c} = 1$ and $\gamma_{\rm s} = 1$,
adding a second antenna at the FC provides a gain of 3.1dB. Adding a
third antenna provides a further gain of 1.34dB, indicating
diminishing returns. To study the benefits of having multiple
antennas, we compare the error exponent in each case with
$\mathcal{E}_{\rm AWGN}(1)$. The multiple antenna gain for the AWGN
case is given by
\begin{equation}
G_{\rm AWGN}(N) \mathop{:=} \frac{\mathcal{E}_{\rm AWGN}(N)}
{\mathcal{E}_{\rm AWGN}(1)} = \frac{N\gamma_{\rm c} + N p_{1}
\gamma_{\rm s} + N} {N \gamma_{\rm c} + p_{1} \gamma_{\rm s} + 1}.
\label{eqn:thm_gain_AWGN}
\end{equation}
It can be seen from (\ref{eqn:thm_gain_AWGN}) that by making $N$
sufficiently large, and $\gamma_{\rm c}$ sufficiently small,
(\ref{eqn:thm_gain_AWGN}) can be made arbitrarily large. In
contrast, it will be seen in Section \ref{ssec:full_CSIS} that when
the channels are fading and known at the sensors, the corresponding
gain expression will be bounded for all parameter values, indicating
limited gains due to antennas.

\section{Performance over Fading Channels}
\label{sec:fading}

Suppose that the elements of $\mathbf{H}$ are non-zero-mean: $h_{nl} = \sqrt{K/(K+1)} + (1/\sqrt{K+1}) h_{nl}^{\rm diff}$,
where the first term is the line-of-sight (LOS) component,
$h_{nl}^{\rm diff}$ is the zero-mean diffuse component, and the
parameter $K$ is the ratio of the LOS power to the power of the
diffuse component, chosen so that the channel satisfies
$E[|h_{nl}|^{2}] = E[|h_{nl}^{\rm diff}|^{2}] = 1$.

In what follows, different cases of channel state information at
the sensors (CSIS) are considered. 

\subsection{No Channel State Information at the Sensors}
\label{ssec:NoCSIS}

When the sensors have no channel knowledge, then the sensor gains
are set to $\boldsymbol{\alpha} = \sqrt{P/L} \mathbf{1}_{L}$ due to
the i.i.d. nature of the channels and to respect the power
constraint in (\ref{eqn:power_constraint}). Substituting in
(\ref{eqn:cov_R}),
\begin{equation}
\mathbf{R} \mathop{:=} \mathbf{R}(\sqrt{P/L} \mathbf{1}_{L}) =
\sigma_{\eta}^{2} P \frac{1}{L} \mathbf{HH}^{H} + \sigma_{\nu}^{2}
\mathbf{I}_{N}. \label{eqn:R_no_CSIS}
\end{equation}
Since the elements of $\mathbf{H}$ are i.i.d., from the weak law of
large numbers,
\begin{equation}
\lim_{L\to\infty} \mathbf{R} = \sigma_{\eta}^{2} P \frac{K}{K+1}
\mathbf{1}_{N\times N} + \frac{\sigma_{\eta}^{2} P +
\sigma_{\nu}^{2} (K + 1)} {K+1} \mathbf{I}_{N},
\label{eqn:R_no_CSIS_asym}
\end{equation}
in probability. Since the right-hand-side of
(\ref{eqn:R_no_CSIS_asym}) is non-singular, $\lim_{L\to\infty}
\mathbf{R}^{-1} = (\lim_{L\to\infty} \mathbf{R})^{-1}$ \cite[Thm.
2.3.4]{Golub}. Using the matrix inversion lemma on
(\ref{eqn:R_no_CSIS_asym}) and substituting into
(\ref{eqn:err_exp_fn}),
\begin{align}
\mathcal{E}_{\rm NoCSIS}(N,K) & = \frac{\theta^{2}}{8} \frac{P(K+1)}
{\sigma_{\eta}^{2}P + \sigma_{\nu}^{2} (K + 1)} \lim_{L\to\infty}
\sum_{n=1}^{N} \left| \frac{1}{L} \sum_{l=1}^{L} h_{nl} \right|^{2}
\nonumber \\ & - \frac{\theta^{2}}{8} \frac{\sigma_{\eta}^{2}
P^{2}K(K+1)} {\left[ \sigma_{\eta}^{2} P + \sigma_{\nu}^{2} (K + 1)
\right] \left[ \sigma_{\eta}^{2}PNK + \sigma_{\eta}^{2}P +
\sigma_{\nu}^{2} (K + 1) \right]} \lim_{L\to\infty} \left|
\frac{1}{L} \sum_{n=1}^{N} \sum_{l=1}^{L} h_{nl} \right|^{2}.
\label{eqn:err_exp_NoCSIS_setup}
\end{align}
Using the weak law of large numbers and
(\ref{eqn:power_constraint}), the error exponent can be expressed in
terms of $\gamma_{\rm c}$ and $\gamma_{\rm s}$ as
\begin{equation}
\mathcal{E}_{\rm NoCSIS}(N,K) \mathop{:=} \frac{1}{8} \frac{NK
\gamma_{\rm c} \gamma_{\rm s}} {\gamma_{\rm c} (NK+1) + \left( p_{1}
\gamma_{\rm s} + 1 \right) \left( K + 1 \right)},
\label{eqn:E_no_CSIS}
\end{equation}
which can be shown to be a monotonically increasing function of $N$,
$K$, $\gamma_{\rm s}$ and $\gamma_{\rm c}$, as expected. For the
single antenna case, using (\ref{eqn:E_AWGN_1}) we have
$\mathcal{E}_{\rm NoCSIS}(1,K) = \mathcal{E}_{\rm AWGN} (1) K/(K + 1)$, 
which is a factor $K/(K+1)$ worse than $\mathcal{E}_{\rm AWGN}(1)$.

As the antennas increase, $\lim_{N\to\infty} \mathcal{E}_{\rm
NoCSIS}(N,K) = \gamma_{\rm s}/8$, which is the same as
$\lim_{N\to\infty} \mathcal{E}_{\rm AWGN}(N)$. That is, so long as
there is some non-zero LOS component, as the number of antennas at
the FC increases, the performance approaches the AWGN performance
even in the absence of CSI at the sensors. Furthermore, it can be
seen that $\lim_{K\to\infty} \mathcal{E}_{\rm NoCSIS}(N,K) =
\mathcal{E}_{\rm AWGN}(N)$, which matches the AWGN result, as
expected.

To characterize the gain due to having multiple antennas at the FC,
we define
\begin{equation}
G_{\rm NoCSIS}(N,K) \mathop{:=} \frac{\mathcal{E}_{\rm NoCSIS}(N,K)}
{\mathcal{E}_{\rm NoCSIS}(1,K)} = \frac{N (K+1) (\gamma_{\rm c} +
p_{1} \gamma_{\rm s} + 1)} {\gamma_{\rm c} (NK+1) + (p_{1}
\gamma_{\rm s} + 1) (K + 1)}. \label{eqn:thm_gain_NoCSIS}
\end{equation}
When the channel noise is large, ($\gamma_{\rm c}\to0$), we have
$G_{\rm NoCSIS}(N,K) = N$ and the gain increases with the number of
antennas at the FC. However, when $\gamma_{\rm c}\to0$, the absolute performance of the system is poor, as can be verified by substituting in (\ref{eqn:E_no_CSIS}). Conversely, when the channel SNR grows, the
maximum gain in (\ref{eqn:thm_gain_NoCSIS}) is given by $(K+1)/K$. This leads
to the conclusion that when the channels between the sensors and the
FC are relatively noise-free, there is little advantage in having
multiple antennas at the FC when $K$ is large. When the channel is
zero-mean ($K=0$), the error exponent in (\ref{eqn:E_no_CSIS}) is
zero for any $N$, indicating that the probability of error does not
decrease exponentially with $L$ for any $N$, confirming results from \cite{Merg06, Liu2007a, banavar09}. However, from
(\ref{eqn:thm_gain_NoCSIS}), it is clear that the {\em gain}
satisfies $\lim_{K\to 0} G_{\rm NoCSIS} (N,K) = N$, which shows that
when the channel is zero-mean, gain in the error exponent due to
antennas is linear and can be made arbitrarily large. We have thus
established the following:
\begin{thm}
For zero-mean channels, with no CSI at the sensors, the error
exponent in (\ref{eqn:E_no_CSIS}) is zero and therefore, the error
probability does not decrease exponentially with $L$ for any number
of antennas, $N$. The antenna gain, defined in
(\ref{eqn:thm_gain_NoCSIS}) satisfies $\lim_{K\to 0} G_{\rm NoCSIS}
(N,K) = N$, implying unlimited gains from multiple antennas for
zero-mean channels when CSI is unavailable at the sensors.
\label{thm:Rayleigh_NoCSIS}
\end{thm}

In what follows, it will be seen that when CSI is available at the
sensors, the antenna gain is bounded over all parameter values for
zero-mean channels.

\subsection{Channel State Information at the Sensors}
\label{ssec:full_CSIS}

We have just seen that when the non-zero-mean channel assumption
does not hold, the incoherent sum of signals at each each antenna
leads to poor performance at the FC, which results in a zero error
exponent. If channel information is available at the sensors, the
sensor gains can be adjusted in such a way that the signals are
combined coherently. It should be noted here that full CSI at the
sensors implies full CSI of the network, $\mathbf{H}$, at the
sensors.  In such a case, $\boldsymbol{\alpha}$ is chosen as a
function of the channels, $\mathbf{H}$.

As a benchmark result for fading channels, the sensor gains are
selected in such a way as to maximize the error exponent of the
system given in (\ref{eqn:err_exp_fn}), subject to the power
constraint in (\ref{eqn:power_constraint}):
\begin{equation}
\boldsymbol{\alpha}_{\rm OPT} = \argmax_{\boldsymbol{\alpha}} \left[
\boldsymbol{\alpha}^{H}\mathbf{H}^{H} \mathbf{R}
(\boldsymbol{\alpha})^{-1} \mathbf{H} \boldsymbol{\alpha} \right]
\quad \text{subject to } \left\|\boldsymbol{\alpha}\right\|^{2} \leq
P, \label{eqn:fullCSIS_alpha_setup}
\end{equation}
to obtain the error exponent in the presence of CSIS,
\begin{equation}
\mathcal{E}_{\rm CSIS}(N) = \lim_{L\to\infty} \frac{\theta^{2}}{8}
\frac{1}{L} \boldsymbol{\alpha}_{\rm OPT}^{H} \mathbf{H}^{H} \left(
\sigma_{\eta}^{2} \mathbf{HD}(\boldsymbol{\alpha}_{\rm OPT})
\mathbf{D}(\boldsymbol{\alpha}_{\rm OPT})^{H} \mathbf{H}^{H} +
\sigma_{\nu}^{2} \mathbf{I}_{N} \right)^{-1} \mathbf{H}
\boldsymbol{\alpha}_{\rm OPT}. \label{eqn:fullCSIS_err_exp_setup}
\end{equation}

The optimization problem in (\ref{eqn:fullCSIS_alpha_setup}) is not
tractable when $N>1$ since $\mathbf{R}(\boldsymbol{\alpha})$ depends
on $\mathbf{H}$ and $\boldsymbol{\alpha}$. In order to assess the
effect of number of antennas, the solution for
(\ref{eqn:fullCSIS_err_exp_setup}) with $N=1$, and two upper bounds
on (\ref{eqn:fullCSIS_err_exp_setup}) are derived for $N>1$.

\subsubsection{Solution for Single Antenna at the FC}
\label{sssec:single_antenna}

When $N=1$, the channel matrix reduces to a column vector, given by
$[h_{1} h_{2} \dots h_{L}]^{T}$, where $h_{i}$ is the channel
between the $i$-th sensor and the FC. The maximization problem in
(\ref{eqn:fullCSIS_alpha_setup}) reduces to
\begin{equation}
\boldsymbol{\alpha}_{\rm OPT} = \argmax_{\boldsymbol{\alpha}}
\frac{\displaystyle{\left| \sum_{i=1}^{L} \alpha_{i} h_{i}
\right|^{2}}} {\displaystyle{\sigma_{\eta}^{2} \sum_{i=1}^{L} \left|
\alpha_{i} \right|^{2} \left| h_{i} \right|^{2} + \sigma_{\nu}^{2}}}
\qquad \text{ subject to } \sum_{i=1}^{L} \left| \alpha_{i}
\right|^{2} \leq P. \label{eqn:opt_prbm_1_setup}
\end{equation}

\noindent A similar problem was formulated in \cite{Wimalajeewa08} and in a distributed
estimation framework in \cite{banavar09, xiao08}. We recognize that
the best value for the phase of the sensor gain is $\angle
\alpha_{l} = - \psi_{l}$ where $\psi_{l} = \angle h_{l}$. Therefore,
we set $\angle \alpha_{l} = - \psi_{l}, \forall l$. We then define
$s \mathop{:=} \sum_{i=1}^{L} \alpha_{i} h_{i}$ and swap the
objective function with the constraint so we can rewrite the
optimization problem as
\begin{align}
\boldsymbol{\alpha}_{OPT} = \argmin_{\{ |\alpha_{i}| \}, s}
\sum_{k=1}^{L} \left| \alpha_{k} \right|^{2} \qquad \text{ subject
to } \quad & \sigma_{\eta}^{2} \sum_{l=1}^{L} \left| \alpha_{l}
\right|^{2} \left| h_{l} \right|^{2} + 1 \leq v_{t} s^{2} \nonumber
\\
& \sum_{l=1}^{L} \left( |\alpha_{l}| |h_{l}| \right) - s = 0,
\label{eqn:opt_prbm_1}
\end{align}
where $v_{t}$ is an auxiliary variable. The optimization problem in
(\ref{eqn:opt_prbm_1}) is now a (convex) second-order-cone problem
\cite{Boyd}. Using the  Karush-Kuhn-Tucker conditions \cite{Boyd},
the optimal solution is given by
\begin{equation}
\alpha_{i} = \sqrt{\frac{P} {\displaystyle{\sum_{l=1}^{L} \left(
\frac{|h_{l}|} {P |h_{l}|^{2} \sigma_{\eta}^{2} + \sigma_{\nu}^{2}}
\right)^{2}}}} \left( \frac{|h_{i}|} {\sigma_{\eta}^{2} P
|h_{i}|^{2} + \sigma_{\nu}^{2}} \right) e^{-j \angle h_{i}}.
\label{eqn:opt_sensor_gains_1}
\end{equation}
The error exponent can be obtained by substituting
(\ref{eqn:opt_sensor_gains_1}) in (\ref{eqn:fullCSIS_err_exp_setup})
with $N=1$:
\begin{equation}
\mathcal{E}_{\rm CSIS}(1) = \lim_{L\to\infty} \frac{\theta^{2}}{8}
\frac{1}{L} \sum_{l=1}^{L} \frac{1} {\sigma_{\eta}^{2} +
\frac{\sigma_{\nu}^{2}} { P |h_{l}|^{2}} } = \frac{\theta^{2}}{8} E
\left[ \frac{1} {\sigma_{\eta}^{2} + \frac{\sigma_{\nu}^{2}} { P
|h_{l}|^{2}} } \right] \label{eqn:opt_err_exp_1}
\end{equation}
from the weak law of large numbers, where the expectation is with
respect to $\{ h_{l} \}$.
As an example, for Rayleigh fading channels
(\ref{eqn:opt_err_exp_1}) yields \cite[\textsection
3.353]{Integrals2007}
\begin{equation}
\mathcal{E}_{\rm CSIS}(1) = \frac{1}{32} \gamma_{\rm s} \left[ 2 -
\frac{p_{1} \gamma_{\rm s} + 1} {\gamma_{\rm c}} \exp \left(
\frac{p_{1} \gamma_{\rm s} + 1} {2 \gamma_{\rm c}} \right) E_{1}
\left( \frac{p_{1} \gamma_{\rm s} + 1} {2 \gamma_{\rm c}} \right)
\right], \label{eqn:opt_err_exp_1_Rayleigh}
\end{equation}
where $E_{1}(\cdot)$ is an exponential integral function \cite[pp.
228]{Abra65}. The expression for $\mathcal{E}_{\rm CSIS}(1)$ is
obtained when the channels between the sensors and the FC are
fading. To compare with the AWGN case, note that
$Px/(\sigma_{\eta}^{2} Px + \sigma_{\nu}^{2})$ in
(\ref{eqn:opt_err_exp_1}) is a concave function of $x$, and from
Jensen's inequality, $\mathcal{E}_{\rm AWGN}(1) \geq
\mathcal{E}_{\rm CSIS}(1)$, as expected.

Since (\ref{eqn:opt_err_exp_1}) is rather complicated, it is
desirable to find a simpler expression as a lower bound to
(\ref{eqn:opt_err_exp_1}). Any choice of
$\|\boldsymbol{\alpha}\|^{2} = P$ will yield such a lower bound,
since $\boldsymbol{\alpha}_{\rm OPT}$ is optimal. Considering
phase-only correction at the sensors, $\alpha_{i} = \sqrt{P/L} \exp
(-j \angle h_{i})$ is substituted in (\ref{eqn:err_exp_fn}) with
$N=1$ to yield the error exponent for phase-only CSIS for $N=1$:
\begin{equation}
\mathcal{E}_{\rm PO}(1) = \lim_{L\to\infty} \frac{\theta^{2}}{8}
\frac{P \displaystyle{\left[ \frac{1}{L} \sum_{l=1}^{L} \left| h_{l}
\right| \right]^{2}}} {\displaystyle{\sigma_{\eta}^{2} P \frac{1}{L}
\sum_{l=1}^{L} |h_{l}|^{2} + \sigma_{\nu}^{2}}}.
\label{eqn:only_phase_bound_setup}
\end{equation}
From the weak law of large numbers, the random sequences in the
numerator and denominator converge separately. However, since the
expression for $\mathcal{E}_{\rm PO}(1)$ is a continuous function of
these sequences, the value of $\mathcal{E}_{\rm PO}(1)$ converges to
\cite[Thm. C.1]{Porat93}
\begin{equation}
\mathcal{E}_{\rm PO}(1) = (E [|h_{l}|])^{2} \mathcal{E}_{\rm
AWGN}(1) \label{eqn:phase_only_bound}
\end{equation}
in probability, since $E[|h_{l}|^{2}] = 1$. The expression in
(\ref{eqn:phase_only_bound}) serves as a lower bound to
$\mathcal{E}_{\rm CSIS}(1)$ as follows:
\begin{equation}
\frac{1} {\zeta} \mathcal{E}_{\rm AWGN}(1) \leq \mathcal{E}_{\rm
CSIS}(1) \leq \mathcal{E}_{\rm AWGN}(1), \label{eqn:po_bound}
\end{equation}
where $\zeta = (E[|h_{l}|])^{-2}$.

\subsubsection{Upper Bound (AWGN channels)} \label{sssec:AWGN_bound}

Since (\ref{eqn:fullCSIS_alpha_setup}) cannot be solved in closed
form when $N>1$, one cannot evaluate the error exponent in
(\ref{eqn:fullCSIS_err_exp_setup}) by substitution as it was done
for $N=1$. Two upper bounds on (\ref{eqn:fullCSIS_err_exp_setup})
will be convenient at this stage. Since the AWGN performance is a
benchmark for fading channels, the error exponent of the system over
AWGN channels is an upper bound on that of fading channels, even in
the case of full CSIS. Therefore, the first upper bound to
(\ref{eqn:fullCSIS_err_exp_setup}) is given in (\ref{eqn:E_AWGN}):
\begin{equation}
\mathcal{E}_{\rm CSIS}(N) \leq \mathcal{E}_{\rm AWGN}(N) =
\frac{1}{8} \frac{N \gamma_{\rm s} \gamma_{\rm c} } {N \gamma_{\rm
c} + p_{1} \gamma_{\rm s} + 1}. \label{eqn:A_bound}
\end{equation}

\subsubsection{Upper Bound (No Sensing Noise)}
\label{sssec:no_obs_noise}

Clearly, (\ref{eqn:fullCSIS_err_exp_setup}) is a monotonically
decreasing function of the sensing noise variance,
$\sigma_{\eta}^{2}$. The second benchmark is obtained by setting
$\sigma_{\eta}^{2} = 0$, which also affects
$\boldsymbol{\alpha}_{\rm OPT}$ in (\ref{eqn:fullCSIS_alpha_setup}),
since $\mathbf{R}(\boldsymbol{\alpha})$ no longer depends on
$\boldsymbol{\alpha}$ when $\sigma_{\eta}^{2} = 0$. Substituting
this in (\ref{eqn:fullCSIS_alpha_setup}), the optimal value of
$\boldsymbol{\alpha}$ when $\sigma_{\eta}^{2} = 0$ is
\begin{equation}
\argmax_{\boldsymbol{\alpha}} \left(\boldsymbol{\alpha}^{H}
\mathbf{H}^{H} \mathbf{H} \boldsymbol{\alpha} \right) \quad
\text{subject to } \left\|\boldsymbol{\alpha}\right\|^{2} \leq P.
\label{eqn:full_CSIS_no_obs_noise_setup}
\end{equation}
The solution to (\ref{eqn:full_CSIS_no_obs_noise_setup}) is the
eigenvector corresponding to the maximum eigenvalue of
$\mathbf{H}^{H}\mathbf{H}$, scaled in a way to satisfy the
constraint with equality. Substituting into
(\ref{eqn:fullCSIS_err_exp_setup}) with $\sigma_{\eta}^{2} = 0$, we
have the second upper bound to $\mathcal{E}_{\rm CSIS}(N)$:
\begin{equation}
B(N,K) = \frac{\theta^{2}} {8} \frac{P} {\sigma_{\nu}^{2}}
\lim_{L\to\infty} \lambda_{\rm max} \left( \frac{1}{L}
\mathbf{H}^{H} \mathbf{H} \right), \label{eqn:B_bound_defn}
\end{equation}
where $\lambda_{\rm max}(\cdot)$ denotes the maximum eigenvalue
function. Using the fact that $\lambda_{\rm max}(\mathbf{H}^{H}
\mathbf{H}) = \lambda_{\rm max}(\mathbf{HH}^{H})$, and that
$\lambda_{\rm max} (\cdot)$ is a continuous function of the matrix
elements \cite[Thm. 8.1.5]{Golub}, one can interchange the limit
with the maximum eigenvalue function \cite[pp. 422, Thm.
C.1]{Porat93} to yield
\begin{equation}
B(N,K) = \frac{\theta^{2}} {8} \frac{P} {\sigma_{\nu}^{2}}
\lambda_{\rm max} \left( \lim_{L\to\infty} \frac{1}{L}
\mathbf{HH}^{H} \right). \label{eqn:B_bound_switch_lim}
\end{equation}
From the weak law of large numbers,
\begin{equation}
\lim_{L\to\infty} \frac{1}{L} \mathbf{HH}^{H} = \frac{K}{K+1}
\mathbf{1}_{N\times N} + \frac{1}{K+1} \mathbf{I}_{N\times N},
\label{eqn:HH}
\end{equation}
in probability, so that with the substitutions $\sigma_{\eta}^{2} =
0$ and $\theta^{2} P/ \sigma_{\nu}^{2} = \gamma_{\rm c}/p_{1}$, we
have the bound:
\begin{equation}
\mathcal{E}_{\rm CSIS} (N) \leq B(N,K) = \frac{1}{8}
\frac{\gamma_{\rm c}} {p_{1}} \frac{NK+1}{K+1}. \label{eqn:B_bound}
\end{equation}
In (\ref{eqn:B_bound}), $B(N,K)$ is an upper bound when there
is sensing noise in the system. When there is no sensing noise, it
is the actual error exponent of the system with full CSIS. Furthermore, $\lim_{K\to\infty} B(N, K) = \lim_{\gamma_{\rm s} \to\infty} \mathcal{E}_{\rm AWGN} (N)$, verifying that  as $K\to\infty$, $B(N,K)$ converges to the AWGN error exponent with no sensing noise. In
addition, if $K=0$, there is no advantage to having multiple
antennas at the FC, for asymptotically large number of sensors,
since the right hand side of (\ref{eqn:B_bound}) is independent of
$N$ in that case.

Since both $\mathcal{E}_{\rm AWGN}(N)$ and $B(N,K)$ are upper bounds
to $\mathcal{E}_{\rm CSIS}(N)$, a combination of the two bounds,
$\min [ \mathcal{E}_{\rm AWGN}(N), B(N,K) ]$, provides a single,
tighter upper bound. Equating the right hand sides of
(\ref{eqn:A_bound}) and (\ref{eqn:B_bound}), it can be shown that
this combined upper bound is given by
\begin{equation}
C(N,K) =
\begin{cases}
\mathcal{E}_{\rm AWGN}(N) & \text{if } \quad \sigma_{\eta}^{2} \geq \frac{N-1}{N(NK+1)} \\
B(N,K) & \text{if } \quad \sigma_{\eta}^{2} \leq \frac{N-1} {N
(NK+1)}
\end{cases}. \label{eqn:CBM_choice}
\end{equation}
Combining the upper and lower bounds,
\begin{equation}
\frac{1} {\zeta} \mathcal{E}_{\rm AWGN}(1) \leq \mathcal{E}_{\rm
CSIS}(1) \leq \mathcal{E}_{\rm CSIS}(N) \leq C(N,K),
\label{eqn:opt_bound}
\end{equation}
obtained from (\ref{eqn:opt_err_exp_1_Rayleigh}),
(\ref{eqn:po_bound}) and (\ref{eqn:CBM_choice}). The bounds in
(\ref{eqn:opt_bound}) will be used to further examine the effect of
$N$ on $\mathcal{E}_{\rm CSIS}(N)$.

The value of $\mathcal{E}_{\rm CSIS}(N)$ from
(\ref{eqn:fullCSIS_err_exp_setup}) is the best achievable
performance for fading channels. Defining the gain due to multiple
antennas as $G_{\rm CSIS}(N) \mathop{:=} \mathcal{E}_{\rm CSIS}(N)/
\mathcal{E}_{\rm CSIS}(1)$, the following theorem can be stated:

\begin{thm}
When the channels have full CSI at the sensors, the gain due to
multiple antennas at the FC can be upper bounded as
\begin{equation}
G_{\rm CSIS}(N) \leq \zeta \frac{\mathcal{E}_{\rm CSIS}(N)}
{\mathcal{E}_{\rm AWGN}(1)} \leq \zeta \min \left[ \frac{N(z+1)}
{Nz+1}, (z+1) \frac{NK + 1} {K + 1} \right],
\label{eqn:thm_gain_opt}
\end{equation}
where $z \mathop{:=} \gamma_{\rm c}/(p_{1} \gamma_{\rm s} + 1)$.
\label{thm:gain_opt}
\end{thm}

\begin{IEEEproof}
The first inequality in (\ref{eqn:thm_gain_opt}) follows from the
first inequality in (\ref{eqn:opt_bound}). The second inequality in
(\ref{eqn:thm_gain_opt}) follows from the last inequality in
(\ref{eqn:opt_bound}) and dividing the terms of
(\ref{eqn:CBM_choice}) by (\ref{eqn:E_AWGN_1}).
\end{IEEEproof}

With $p_{1} = 0.5$, $K = 1$, $\gamma_{\rm c} = 1$ and $\gamma_{\rm
s} = 1$, for $N = 2$, $G_{\rm CSIS}(2) \leq 1.4286 \zeta $. For $N =
3$, $G_{\rm CSIS}(3) \leq 1.6667 \zeta$ and for $N = 4$, $G_{\rm
CSIS}(4) \leq 1.8182 \zeta$. These results indicate that there is diminishing returns in the multiple antenna gain.

\begin{corollary}
$G_{\rm CSIS}(N)$ can be bounded by an expression depending on $N$
and $K$ only:
\begin{equation}
G_{\rm CSIS}(N) \leq \zeta \frac{N^{2}K + 2N - 1} {N (K+1)}
\label{eqn:cor_gain_opt}
\end{equation}
\label{cor:gain_opt}
\end{corollary}

\begin{IEEEproof}
The first argument of the $\min[\cdot,\cdot]$ function of the right
hand side of (\ref{eqn:thm_gain_opt}) is a decreasing function in
$z$ and the second argument is an increasing function in $z$.
Therefore, when the arguments are equal for fixed values of $N$ and
$K$, the maximum value of the $\min[\cdot,\cdot]$ function is
obtained. This occurs when $z = N^{-1}(NK+1)^{-1}(N-1)$, allowing us
to upper bound the $\min[\cdot,\cdot]$ function by the value in
(\ref{eqn:cor_gain_opt}).
\end{IEEEproof}

\begin{corollary}
When the channels have zero-mean, the maximum gain due to having
multiple antennas at the FC is bounded by a constant independent of
$N$ and only dependent on $\zeta = (E[|h_{l}|])^{-2}$:
\begin{equation}
G_{\rm CSIS}(N) \leq 2 \zeta. \label{eqn:gain_rayleigh_cor}
\end{equation}
 \label{cor:gain_Rayleigh}
\end{corollary}

\begin{IEEEproof}
Substituting $K = 0$, it is clear that (\ref{eqn:cor_gain_opt}) is
monotonically increasing in $N$. Taking the limit as $N \to \infty$
yields the proof.
\end{IEEEproof}

As an example, in the case of Rayleigh fading, when full channel
information is available at the sensors, the maximum gain that can
be obtained by adding any number of antennas at the FC for any
channel or sensing SNR is at most $2 \zeta = 8/\pi$, which is less
than 3.

The results in
(\ref{eqn:thm_gain_opt})-(\ref{eqn:gain_rayleigh_cor}) have been
derived for the case of iid sensing noise. We now address the
correlated sensing noise case. To this end, we define
$\mathbf{R}_{\eta}$ as the $L\times L$ covariance matrix of the sensing noise samples, 
$\{\eta_{l}\}_{l=1}^{L}$.
\begin{thm}
\label{thm:corr_obs_noise} Suppose that the sensing noise samples
are correlated and let $\lambda_{\rm min}$ be the minimum eigenvalue
of $\mathbf{R}_{\eta}$. The gain due to multiple antennas in
(\ref{eqn:thm_gain_opt}) holds with the change $z = \gamma_{\rm c}/
(p_{1} \tilde{\gamma}_{\rm s} + 1)$, where $\tilde{\gamma}_{\rm s}
\mathop{:=} \theta^{2}/\lambda_{\rm min}$.
\end{thm}

\begin{IEEEproof}
The proof is shown in Appendix \ref{app:proof_corr_obs_noise}.
\end{IEEEproof}

Theorem \ref{thm:corr_obs_noise} shows that any full-rank sensing
noise covariance matrix changes the conclusion in
(\ref{eqn:thm_gain_opt}) only through a redefinition of $z$. By
maximizing over $z$, the same upper-bound in
(\ref{eqn:cor_gain_opt}) is obtained, and for zero-mean channels,
the bound in (\ref{eqn:gain_rayleigh_cor}) remains valid. This shows
that the bounds in (\ref{eqn:cor_gain_opt}) and
(\ref{eqn:gain_rayleigh_cor}) are general, and hold even when the
iid condition is relaxed to any arbitrary full-rank covariance matrix, $\mathbf{R}_{\eta}$. The gain due to
adding multiple antennas is still upper-bounded by a factor of
$2\zeta$, for zero-mean channels, when there is full CSI at the
sensors.

\subsection{Phase-only CSIS}
\label{ssec:partial_CSIS}

One simplification to the full CSIS case is to provide only channel
phase information to the sensors. For the single antenna case, and
when the channels between the sensors and the FC have zero-mean, the
phase-only results have been presented in
(\ref{eqn:phase_only_bound}) and (\ref{eqn:po_bound}). What follows
is an extension of those results to the multiple antenna case when
$K=0$.

Since there is only phase information at the sensors, the amplitudes
of the sensor gains are selected such that $|\alpha_{l}| =
\sqrt{P/L}, \forall l$, so that $\mathbf{D} (\boldsymbol{\alpha})
\mathbf{D} (\boldsymbol{\alpha})^{H} = (P/L) \mathbf{I}_{L}$ and
$\mathbf{R}(\boldsymbol{\alpha})$ is given by (\ref{eqn:R_no_CSIS}).

With phase-only information, one can constrain $|\alpha_{i}|$ to be constant to reformulate (\ref{eqn:fullCSIS_alpha_setup}) as the following:
\begin{equation}
\boldsymbol{\alpha}_{\rm PO} = \argmax_{\boldsymbol{\alpha}} \boldsymbol{\alpha}^{H} \mathbf{H}^{H} \mathbf{H} \boldsymbol{\alpha} \quad \text{subject to} \quad |\alpha_{i}|^{2} = \frac{P}{L}, i = 1, 2, \dots, L. 
\label{eqn:po_setup}
\end{equation} 
In Section \ref{sssec:sdr}, a semidefinite relaxation approach will be presented to solve (\ref{eqn:po_setup}). 

\subsection{Asymptotically large sensors {\em and} antennas}
\label{ssec:inf_ant_sensors}

When CSIS is available, (\ref{eqn:thm_gain_opt} -
\ref{eqn:gain_rayleigh_cor}) shows that only limited multiple
antenna gains are available. It is interesting to see whether such
limits would still be present if $N \to \infty$ {\em simultaneously}
with $L$. A similar problem was considered, but in the context of CDMA transmissions in \cite{Jayaweera07}. Note that this will in general yield results different
than first sending $L \to \infty$ and then $N \to \infty$ as was
done in Section \ref{ssec:full_CSIS}. Such a situation can be
interpreted as a case where a group of sensors is transmitting to
another group, functioning as a virtual antenna array
\cite{Sendonaris03a}. For such a system the scaling laws when $L$
and $N$ simultaneously increase \cite[pp. 7]{Verdu04}, in such a way
that
\begin{equation}
\lim_{L,N \to \infty} \frac{L}{N} = \beta, \label{eqn:LNratio}
\end{equation}
are of interest. It should be noted that in spite of scaling the
number of sensors and antennas, the power constraint is still
maintained.

In this case, the error exponent is redefined as
\begin{equation}
\mathcal{E}^{\infty} (\beta) = \lim_{L,N\to\infty} - \frac{1}{L}
\log P_{e| \mathbf{H}}(N), \label{eqn:err_exp_inf}
\end{equation}
with (\ref{eqn:LNratio}) satisfied. Similar to the upper bounds in
(\ref{eqn:A_bound}) and (\ref{eqn:B_bound}), upper bounds on
(\ref{eqn:err_exp_inf}) are now derived. For the AWGN case,
\begin{equation}
\mathcal{E}^{\infty} (\beta) \leq \mathcal{E}_{\rm AWGN}^{\infty}
\mathop{:=} \lim_{L,N \to\infty} \mathcal{E}_{\rm AWGN} (N) =
\lim_{L,N \to\infty} \frac{1}{8} \frac{N \gamma_{\rm s} \gamma_{\rm
c}} {N \gamma_{\rm c} + p_{1} \gamma_{\rm s} + 1} = \frac{1}{8}
\gamma_{\rm s}. \label{eqn:A_bound_beta}
\end{equation}
When there is no sensing noise, with $\sigma_{\eta}^{2} = 0$, the
second bound can be calculated as
\begin{equation}
\mathcal{E}^{\infty} (\beta) \leq B^{\infty}(\beta) \mathop{:=}
\lim_{L,N \to \infty} \frac{\theta^{2}}{8} \frac{P}
{\sigma_{\nu}^{2}} \lambda_{\rm max} \left( \frac{1}{L}
\mathbf{H}^{H} \mathbf{H} \right).
\label{eqn:err_exp_no_obs_noise_beta}
\end{equation}
For fading channels with $K>0$, it can be shown that the error
exponent in (\ref{eqn:err_exp_no_obs_noise_beta}) goes to infinity.
Therefore, with any line-of-sight (LOS) and no sensing noise,
increasing the number of sensors and the number of antennas to
infinity provides very good performance. When $K=0$, the
Mar\u{c}enko-Pastur Law \cite[pp. 56]{Verdu04} provides an empirical
distribution of the eigenvalues of $N^{-1} \mathbf{H}^{H}
\mathbf{H}$. From \cite{Yin88, Bai93}, the maximum eigenvalue of
$N^{-1} \mathbf{H}^{H} \mathbf{H}$ is shown to converge in such a
way that
\begin{equation}
\lim_{L,N \to\infty} \lambda_{\rm max} \left[ \left(
\frac{1}{\sqrt{N}} \mathbf{H} \right)^{H} \left( \frac{1}{\sqrt{N}}
\mathbf{H} \right) \right] = \frac{\left( 1 + \sqrt{\beta}
\right)^{2}} {\beta}, \label{eqn:marcenko_pastur_result}
\end{equation}
in probability, which yields
\begin{equation}
B^{\infty}(\beta) = \frac{1}{8} \frac{\gamma_{\rm c}} {p_{1}}
\frac{(1 + \sqrt{\beta})^{2}} {\beta}, \label{eqn:B_bound_beta}
\end{equation}
which is the optimum performance of the system in the absence of
sensing noise. Similar to (\ref{eqn:CBM_choice}), the minimum of
(\ref{eqn:A_bound_beta}) and (\ref{eqn:B_bound_beta}) yields
\begin{equation}
\mathcal{E}^{\infty} (\beta) \leq \min \left[ \mathcal{E}_{\rm
AWGN}^{\infty}, B^{\infty}(\beta) \right] =
\begin{cases}
\frac{1} {8} \gamma_{\rm s} & \text{if } \quad P
\sigma_{\eta}^{2} \geq \frac{\beta} {(1 + \sqrt{\beta})^{2}} \\
\frac{1}{8} \frac{\gamma_{\rm c}} {p_{1}} \frac{(1 +
\sqrt{\beta})^{2}} {\beta} & \text{if } \quad P \sigma_{\eta}^{2}
\leq \frac{\beta} {(1 + \sqrt{\beta})^{2}}
\end{cases}.
\label{eqn:C_inf}
\end{equation}
The gain due to antennas is expressed in terms of the ratio $\beta$
in (\ref{eqn:LNratio}) as $G^{\infty}(\beta) \mathop{:=}
\mathcal{E}^{\infty}(\beta) / \mathcal{E}_{\rm CSIS}(1)$. Using the
bounds, we have the following:
\begin{thm}
With asymptotically large number of sensors and antennas, the gain
due to having multiple antennas at the FC is bounded by
\begin{equation}
G^{\infty} (\beta) \leq \zeta \left( 1 + \frac{\left( 1 +
\sqrt{\beta} \right)^{2}} {\beta} \right). \label{eqn:thm_gain_inf}
\end{equation}
\label{thm:gain_inf}
\end{thm}

\begin{IEEEproof}
The relationship between $\mathcal{E}_{\rm AWGN}(1)$ and
$\mathcal{E}_{\rm CSIS}(1)$ from (\ref{eqn:po_bound}) provides a
lower bound on $\mathcal{E}_{\rm CSIS}(1)$, and consequently an
upper bound on $G^{\infty} (\beta)$, to yield the first inequality
in (\ref{eqn:gain_inf_bound}) below. The expression in
(\ref{eqn:C_inf}) provides an upper bound on $\mathcal{E}^{\infty}
(\beta)$, and dividing by (\ref{eqn:E_AWGN_1}) yields the second
inequality in
\begin{equation}
G^{\infty} (\beta) \leq \zeta \frac{\mathcal{E}^{\infty} (\beta)}
{\mathcal{E}_{\rm AWGN}(1)} \leq \zeta \min \left[ 1 + \frac{1}{w},
(1 + w) \frac{(1 + \sqrt{\beta})^{2}} {\beta} \right],
\label{eqn:gain_inf_bound}
\end{equation}
where $w \mathop{:=} \gamma_{\rm c}/(p_{1} \gamma_{\rm s} + 1)$. The
first argument in the $\min [\cdot, \cdot]$ function decreases as
$w$ increases, while the second argument is an increasing function
of $w$. Therefore, the $\min [\cdot, \cdot]$ function is maximized
when arguments of the $\min [\cdot, \cdot]$ function are equal for a
fixed value of $\beta$. This result is obtained when $w = (1 +
\sqrt{\beta})^{-2} \beta$, to yield (\ref{eqn:thm_gain_inf}) and the
proof.
\end{IEEEproof}

To interpret (\ref{eqn:thm_gain_inf}), cases corresponding to three
values of $\beta$, are considered:
\begin{enumerate}[(i)]
 \item $\beta \ll 1 \text{ } (N \text{ scales faster than } L)$:
 When the number of antennas increases at a faster rate than the
 number of sensors, it can be seen that $B^{\infty}(\beta)$ is
 large. When there is no sensing noise, the performance obtained is
 exactly $B^{\infty}(\beta)$ as seen in
 (\ref{eqn:B_bound_beta}). In this case, arbitrarily large gains
 are achievable. In case there is sensing noise in the system,
 $\mathcal{E}_{\rm AWGN}^{\infty}$ and $B^{\infty}(\beta)$ become
 bounds, and the gain is bounded as shown in (\ref{eqn:thm_gain_inf}).
 As $\beta \to 0$ in this case, the bound goes to infinity, which
 indicates that there could be large gains possible.

 \item $\beta = 1 \text{ } (N \text{ scales as fast as } L)$: The
 number of antennas at the FC and the number of sensors scale at the
 same rate, the maximum possible gain can be calculated from
 (\ref{eqn:thm_gain_inf}) to yield $G^{\infty} (1) \leq 5 \zeta$.

 \item $\beta \gg 1 \text{ } (N \text{ scales slower than } L)$:
 When the number of sensors
 scales much faster than the number of antennas at the FC, it
 resembles the previous setting where $L\to\infty$, first, and
 $N$ was scaled. Not
 surprisingly, when $\beta$ is large in this case, $G^{\infty}
 (\beta) \leq 2 \zeta$, same as in Section \ref{ssec:full_CSIS}.
\end{enumerate}
It should be noted here that in cases (ii) and (iii), where both the
number of sensors and antennas are scaled to infinity
simultaneously, only limited gain is achievable, when the sensors
have complete channel knowledge.

\begin{table}[tb]
\begin{centering}

\begin{tabular}{|c|c|c||c|}
\hline
 & $G_{\rm NoCSIS} (N,K)$ from (\ref{eqn:thm_gain_NoCSIS})& $G_{\rm CSIS}(N,K)$ from (\ref{eqn:cor_gain_opt}) & $G^{\infty} (\beta)$ from (\ref{eqn:thm_gain_inf}) \\
\hline
$K > 0$ & $O(N)$ when $\gamma_{\rm c} = 0; O(1)$ when $\gamma_{\rm c} > 0$ & $O(N)$ & Undefined \\
\hline
$K \to 0$ & $O(N)$ & $O(1)$ & $O(\beta^{-1})$ as $\beta \to 0; O(1)$ as $\beta \to \infty$  \\
\hline
\end{tabular}
 \caption{Order of gain due to multiple antennas at the FC for large number of sensors, $L$.}
 \label{table:summary_results}
\end{centering}
\end{table}

In Table \ref{table:summary_results} we summarize the rate at which
the gain due to number of antennas increases, both when CSI is
available and unavailable at the sensor side. Recalling that the
gain is defined in terms of the ratio of error exponents relative to
the single antenna case, all the results in the table apply when $L$
is large, which is a major distinguishing factor between this study
and standard analysis of multi-antenna systems. It is seen that when
$K>0$ the gain in error exponent grows like $O(N)$ depending on
whether CSIS is available and whether $\gamma_{\rm c} = 0$. More
interestingly, when the channel is zero-mean ($K \to 0$), adding
antennas improves the error exponent linearly when CSIS is not
available. In stark contrast, when CSIS is available, the gain is
bounded ($O(1)$) by $2 \zeta$. Finally, the column on the right of
Table \ref{table:summary_results} illustrates how the gain depends
on the ratio $\beta = L/N$ as both $N$ and $L$ increase. The error
exponents for $K>0$ are infinite, yielding an undefined gain. For
zero-mean channels, the dependence on $\beta$ indicates an
increasing gain when $\beta$ is small ($L \ll N$), and bounded gain
when $\beta$ is large ($L \gg N$).

\subsection{Realizable Schemes}
\label{ssec:prac_scheme}

So far, we have provided bounds on the achievable gains due to
antennas when CSI is available at the sensors, without providing a
realizable scheme. This is because the calculation of
$\boldsymbol{\alpha}_{\rm OPT}$ in (\ref{eqn:fullCSIS_alpha_setup})
in closed form is intractable. Moreover, it is not clear how
$\boldsymbol{\alpha}$ should be chosen as a function of $\mathbf{H}$
when $N>1$ to achieve a multiple-antenna gain. This is because each
sensor sees $N$ channel coefficients, corresponding to $N$ antennas,
and each channel coefficient has a different phase making the
choices of $\angle \alpha_{i}$ non-trivial. We now present two
sub-optimal schemes for the full CSIS case that are shown to provide
gains over the single antenna case.

\subsubsection{Method I: Optimizing Gains to Match the Best Antenna}
\label{sssec:prac_antenna_sel}

In this method, the sensor gains, $\boldsymbol{\alpha}$, are
selected in order to target the best receive antenna. However, the
received signals at all of the other antennas are also combined at
the FC, which uses the detection rule defined in
(\ref{eqn:choose_H1}).

Since $L$ is finite for any practical scheme,
(\ref{eqn:opt_sensor_gains_1}) will be used to select
$\boldsymbol{\alpha}$ and (\ref{eqn:opt_err_exp_1}) without the
limit can be used to assess which antenna has the ``best'' channel
coefficients. Therefore, using the channels from the sensors to all
of the receive antennas,
\begin{equation}
n^{*} = \argmax_{n} \frac{\theta^{2}}{8} \frac{1}{L} \sum_{l=1}^{L}
\frac{1} {\sigma_{\eta}^{2} + \frac{\sigma_{\nu}^{2}} {P
|h_{nl}|^{2}}}, \label{eqn:err_exp_as_first}
\end{equation}
is calculated and the sensor gains are set to
(\ref{eqn:opt_sensor_gains_1}) computed for the channels $\{h_{n^{*}
i}\}_{i=1}^{L}$.
The FC then uses all of the receive antennas for detection using
(\ref{eqn:choose_H1}). Since there are multiple antennas at the FC,
for any realization of the channels between the sensors and the FC,
the error exponent of this scheme is at least as good as the single
antenna case.

Such an approach requires the calculation of
(\ref{eqn:err_exp_as_first}) and the corresponding
$\boldsymbol{\alpha}$ from (\ref{eqn:opt_sensor_gains_1}). Since
these calculations require the complete knowledge of $\mathbf{H}$,
they can be calculated at the FC, and fed back to the sensors.

\subsubsection{Method II: Maximum Singular Value of the Channel Matrix}
\label{sssec:prac_max_sing_value}

It was shown in Section \ref{sssec:no_obs_noise} that when
$\sigma_{\eta}^{2} = 0$, the bound obtained in (\ref{eqn:B_bound})
is achievable. In this method, the values of $\boldsymbol{\alpha}$
are selected as though there is no sensing noise. The sensor gains,
$\boldsymbol{\alpha}$, are selected in such a way that they are a
scaled version of the eigenvector corresponding to $\lambda_{\rm
max} \left( \mathbf{H}^{H} \mathbf{H} \right)$, such that
$\|\boldsymbol{\alpha}\|^2 = P$. In most practical cases, sensing
noise is non-zero, and therefore, this method is sub-optimal.
Similar to Method I, $\boldsymbol{\alpha}$ can be calculated at the
FC and fed back to the sensors.

\subsubsection{Hybrid of Methods I and II}
\label{sssec:hybrid_scheme}

Since Method II is tuned to perform optimally when there is no
sensing noise, it outperforms Method I when the
sensing SNR, $\gamma_{\rm s}$, is high. As the sensing SNR reduces,
Method I begins to outperform Method II. These observations are
illustrated and elaborated on in the simulations section (Section
\ref{sec:simulations}, Figure
\ref{Fig:prac_N5_N50_p15_SNR01_P_T1_20}).

Since one of the schemes performs better than the other based on the
value of $\gamma_{\rm s}$, a hybrid scheme can be used: Method I for
low values of $\gamma_{\rm s}$, and Method II for high values. The
exact value where the cross-over occurs depends on the parameters of
the system, and can determined empirically. An example is shown in
the simulation section in Figure
\ref{Fig:prac_N5_N50_p15_SNR01_P_T1_20}, where it is also argued
that an underestimation of the value of $\gamma_{\rm s}$ is
tolerable, while an overestimation is not.

\subsubsection{Semidefinite Relaxation}
\label{sssec:sdr}

Following \cite{xiao08, Luo2010} a semidefinite relaxation of the problem in (\ref{eqn:po_setup}) is obtained as follows:
\begin{align}
\mathbf{X}_{\rm PO} = \argmax_{\mathbf{X}} \text{trace}(\mathbf{H}^{H} \mathbf{H} \mathbf{X}) \quad \text{subject to} \quad & \mathbf{X} \succeq 0, \nonumber \\
& \mathbf{X}_{ii} = \frac{P}{L}, i = 1, 2, \dots, L,
\label{eqn:po_sdr}
\end{align} 
where $\mathbf{X}$ is an $L\times L$ matrix. If $\mathbf{X}$ has a rank-1 decomposition, $\mathbf{X} \mathop{:=} \boldsymbol{\alpha \alpha}^{H}$, then $\boldsymbol{\alpha}$ is a solution to (\ref{eqn:po_setup}) \cite{xiao08, Luo2010}. In the more likely case where $\mathbf{X}$ does not have rank-1, then an approximation to the solution of (\ref{eqn:po_setup}) is obtained by choosing $\boldsymbol{\alpha}$ as the vector consisting of the phases of the eigenvector corresponding to the maximum eigenvalue of $\mathbf{X}$. The semidefinite relaxation in (\ref{eqn:po_sdr}) causes a loss of upto a factor of $\pi/4$ in the final answer of (\ref{eqn:po_setup}) \cite{Luo2010}. The phases of eigenvector corresponding to the maximum eigenvalue of $\mathbf{X}_{\rm PO}$ are extracted to constitute a possible set of values of $\boldsymbol{\alpha}$. In order to obtain the solution to the SDR problem, an eigenvalue decomposition of $\mathbf{X}_{\rm OPT}$ is required, which is an $O(L^{3})$ operation \cite{Golub}. It is argued with the help of simulations (Figure \ref{Fig:sdr_hybrid}) that the SDR outperforms the hybrid scheme when $\gamma_{\rm s}$ is small, at the expense of increased complexity. 

\section{Simulation Results}
\label{sec:simulations}

\begin{figure}[tb]
\begin{minipage}[b]{1.0\linewidth}
  \centering
  \centerline{\epsfig{figure=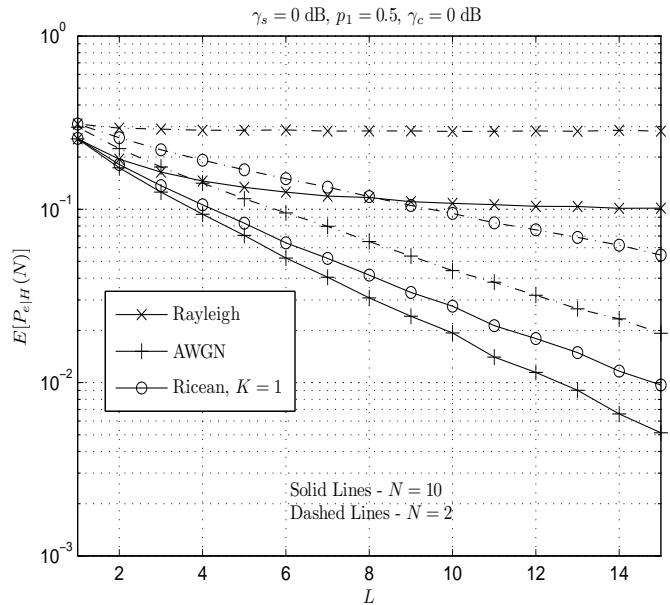,width=10cm,height=8.5cm}}
\end{minipage}
\caption{Monte-Carlo Simulation: $E[P_{e|\mathbf{H}}(N)]$ for AWGN
channels, Rayleigh fading channels and Ricean channels with no
CSIS.} \label{Fig:AWGN_Rayleigh_Ricean_no_CSIS_N2_10_Pe_L1to15}
\end{figure}

\begin{figure}[tb]
\begin{minipage}[b]{1.0\linewidth}
  \centering
  \centerline{\epsfig{figure=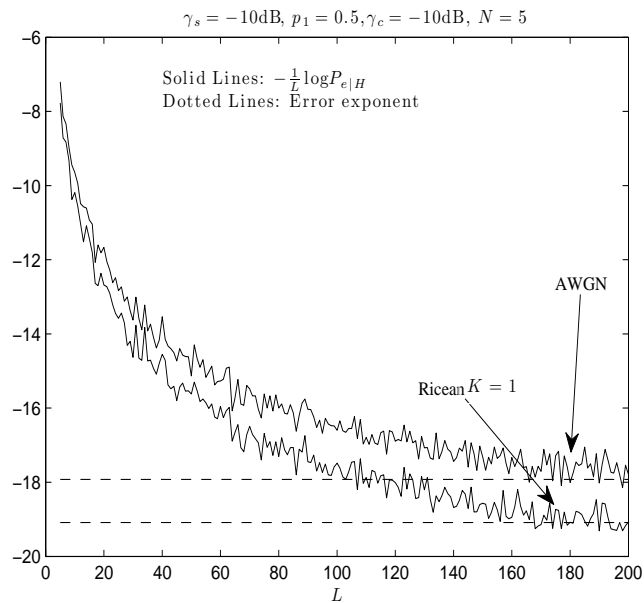,width=10cm,height=8.5cm}}
\end{minipage}
\caption{Monte-Carlo simulation - Error exponent for AWGN and Ricean
Fading channels.} \label{Fig:mc_e}
\end{figure}

\begin{figure}[tb]
\begin{minipage}[b]{1.0\linewidth}
  \centering
  \centerline{\epsfig{figure=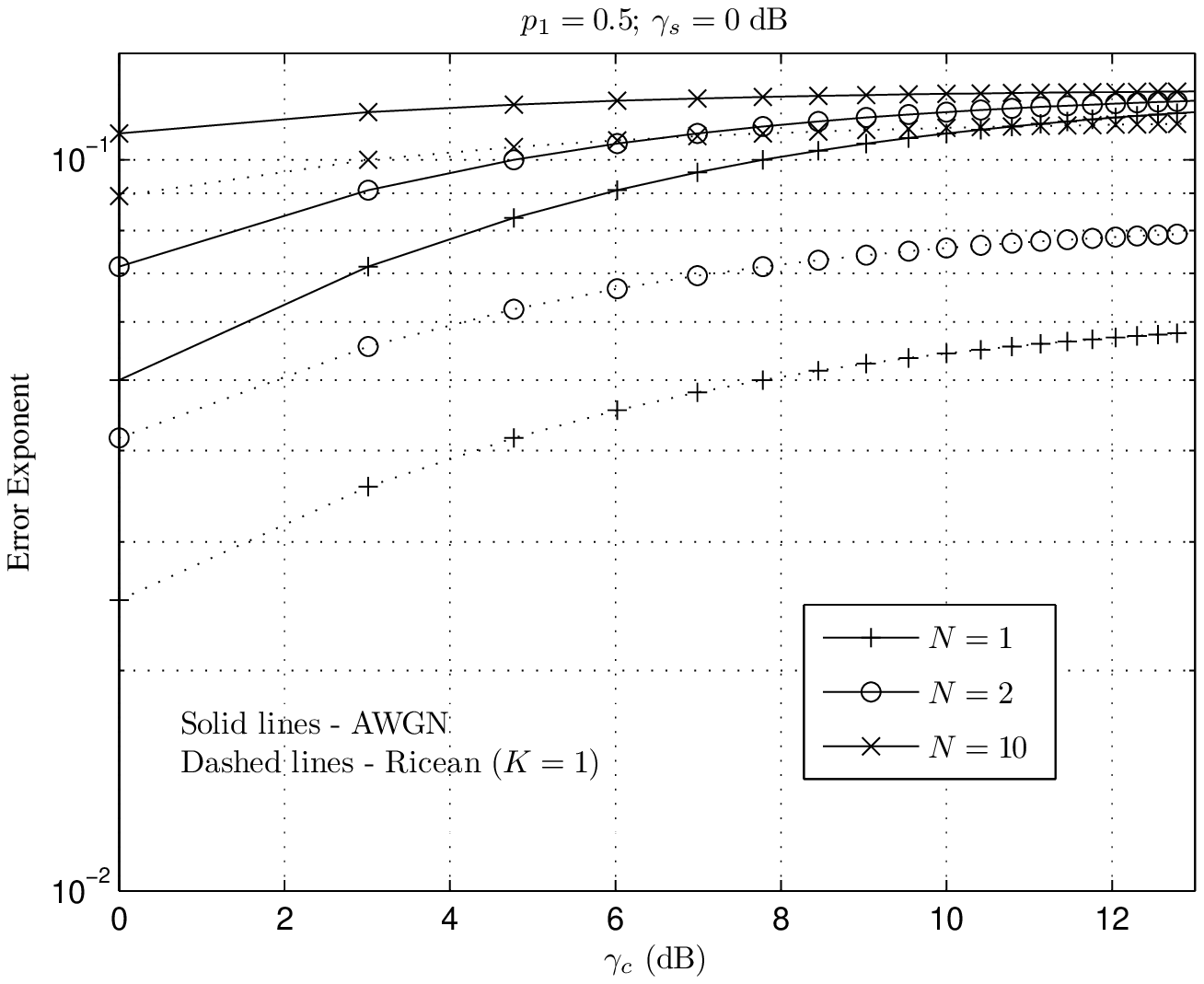,width=10cm,height=8.5cm}}
\end{minipage}
\caption{Error exponent vs $\gamma_{\rm s}$ for $N = 1,2,10$ for
AWGN channels and Ricean channels and no CSIS.}
\label{Fig:AWGN_Ricean_N1_N2_N10_PT1_p15_SNR_0_20}
\end{figure}

The theoretical results obtained are verified using simulations. The
channels are generated as complex Gaussian (Rayleigh or Ricean) for
the purposes of simulation, even though the results only depend on
the first and second order moments of the channels.

In Figure \ref{Fig:AWGN_Rayleigh_Ricean_no_CSIS_N2_10_Pe_L1to15}, it
is verified that increasing the number of sensors improves the
performance except when the channels are Rayleigh fading and there
is no CSIS. Since the error exponent is zero for the Rayleigh fading
case with no CSIS, the asymptotic average probability of error is
computed and plotted.
The Ricean case outperforms the Rayleigh fading case, and the AWGN
channels provide the best performance. It can also be seen that the
decay in probability of error is exponential in $L$, when the
channels between the sensors and the FC are AWGN or Ricean fading.
The decay is slower than exponential when the channels are Rayleigh
fading. This confirms the observations in Section \ref{ssec:NoCSIS}.
In all cases, the performance improves as the number of antennas
increases.

In Figure \ref{Fig:mc_e}, the expression of error exponent is
compared against the value of $L^{-1} \log P_{e|\mathbf{H}} (5)$ for
increasing $L$, with AWGN channels and Ricean fading channels
between the sensors and the FC. It can be seen that fewer than 200
sensors are required for the asymptotic results to hold. Therefore,
in subsequent simulations, $L = 200$ sensors have been used.

The effect of increasing the number of antennas on the error
exponent for the AWGN case and Ricean fading case with no CSIS is
seen in Figure \ref{Fig:AWGN_Ricean_N1_N2_N10_PT1_p15_SNR_0_20}. As
expected, increasing $\gamma_{\rm c}$ improves performance and there
is an improvement in performance as the number of antennas at the FC
increases. As predicted in Section \ref{ssec:NoCSIS}, with an
increase in $N$, the performance of $\mathcal{E}_{\rm AWGN}(N)$ and
$\mathcal{E}_{\rm NoCSIS}(N,K)$ get closer to each other. There is a
large performance gain between the $N=1$ case and the $N=2$ case,
and almost the same gain between the $N=2$ case and the $N=10$ case,
indicating diminishing returns, corroborating the results in Section
\ref{sec:AWGN}.

\begin{figure}[tb]
\begin{minipage}[b]{1.0\linewidth}
  \centering
  \centerline{\epsfig{figure=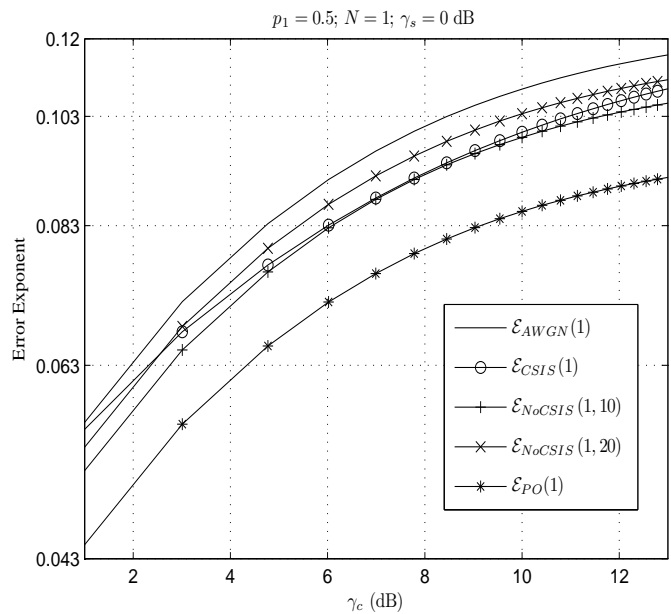,width=10cm,height=8.5cm}}
\end{minipage}
\caption{Optimal Rayleigh performance, AWGN performance and Ricean
no CSIS performance with one antenna at the FC.}
\label{Fig:AWGN_OPT_Ricean_K10_K20_N1_SNR1_p15_P_T2_20}
\end{figure}

In Figure \ref{Fig:AWGN_OPT_Ricean_K10_K20_N1_SNR1_p15_P_T2_20}, the
error exponent is evaluated when there is a single antenna at the
FC. The cases of AWGN channels, Ricean channels with no CSIS,
Rayleigh fading channels with full CSIS  and Rayleigh fading
channels with phase-only CSIS are compared in Figure
\ref{Fig:AWGN_OPT_Ricean_K10_K20_N1_SNR1_p15_P_T2_20}. It is seen
that the AWGN performance is the best, and when the Ricean channels
have larger line of sight, the performance improves, as expected. In
fact, by increasing the amount of LOS, the no-CSIS Ricean case
performs better than the full CSIS Rayleigh channel case, when
$\gamma_{\rm c}$ is large. The performance of the Ricean no CSIS
case is a constant factor $K/(K+1)$ worse than the AWGN case,
corroborating the result of $\mathcal{E}_{\rm NoCSIS} (1,K)$. Similarly, the
performance of the phase-only CSIS case confirms the result in
(\ref{eqn:po_bound}). For Rayleigh fading channels, the phase-only
CSIS case performs a constant $\pi/4$ worse than the AWGN case.

\begin{figure}[tb]
\begin{minipage}[b]{1.0\linewidth}
  \centering
  \centerline{\epsfig{figure=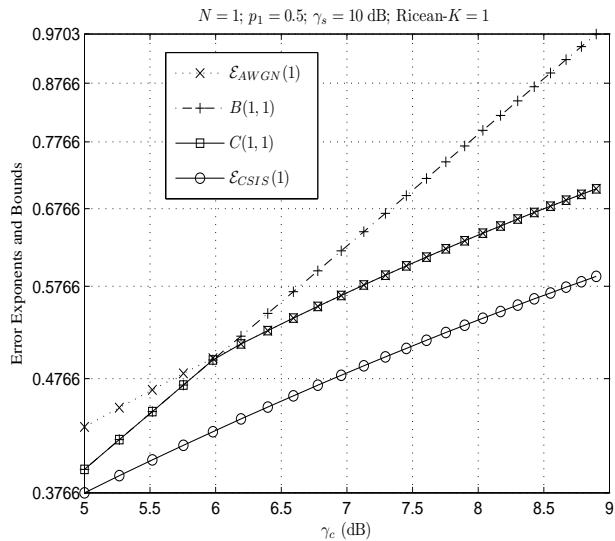,width=9cm,height=7.5cm}}
\end{minipage}
\caption{For a single antenna, optimal performance and performance bounds.}
\label{Fig:A_B_E_N1_obs_SNR10_P_T3to6}
\end{figure}

For the case of full CSIS, but with multiple antennas at the FC,
bounds were derived on the error exponent of the system in Section
\ref{sssec:AWGN_bound} and Section \ref{sssec:no_obs_noise}, and
combined to provide a single bound in (\ref{eqn:CBM_choice}). The
value of $\mathcal{E}_{\rm CSIS}(1)$ is set as a lower-bound on
$\mathcal{E}_{\rm CSIS}(N)$. In Figure
\ref{Fig:A_B_E_N1_obs_SNR10_P_T3to6}, with $N=1$, the upper bound
can be seen to be about 0.76 dB (in terms of error exponent) away
from the actual value at $\gamma_{\rm c} = 8$ dB. For small values
of $\gamma_{\rm c}$, the AWGN bound is better, and as $\gamma_{\rm
c}$ increases, the bound with the no sensing noise assumption is
better, as expected. 

\begin{figure}[tb]
\begin{minipage}[b]{1.0\linewidth}
  \centering
  \centerline{\epsfig{figure=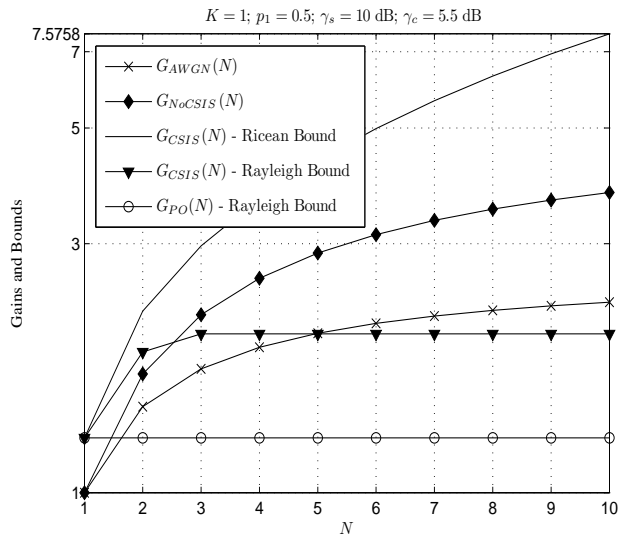,width=9cm,height=7.5cm}}
\end{minipage}
\caption{Comparison of antenna gains vs $N$.}
\label{Fig:A_C_RiceNoCSIS_p15_P_T001_SNR05_K1_N1_100}
\end{figure}

Figure \ref{Fig:A_C_RiceNoCSIS_p15_P_T001_SNR05_K1_N1_100} shows the
effect of increasing the number of antennas at the FC on the antenna
gains of the different systems. Also, for the cases of partial CSIS
and full CSIS, the upper bounds on the antenna gains are plotted.
The actual error exponent for the AWGN case is larger than for the
Ricean no-CSIS case. However, as seen in Figure
\ref{Fig:A_C_RiceNoCSIS_p15_P_T001_SNR05_K1_N1_100}, the gain for
the Ricean no-CSIS case is larger than the gain for the AWGN channel
case. The bound on the Ricean CSIS antenna gain grows rapidly with
$N$, as predicted by (\ref{eqn:cor_gain_opt}). The maximum gains
possible for the Rayleigh CSIS case and the Rayleigh no CSIS cases
are also plotted. These results indicate that with full CSIS, there
is not much to be gained by adding antennas at the FC, corroborating
our results in Section \ref{ssec:full_CSIS}.

\begin{figure}[tb]
\begin{minipage}[b]{1.0\linewidth}
  \centering
  \centerline{\epsfig{figure=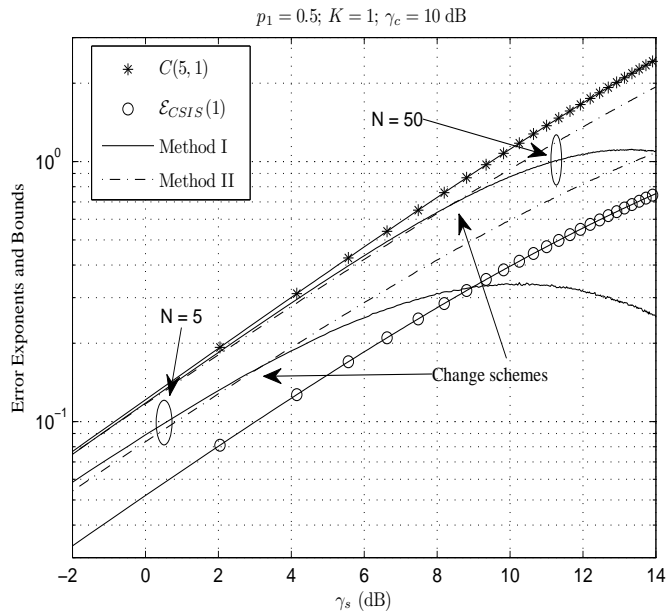,width=10cm,height=8.5cm}}
\end{minipage}
\caption{Practical Schemes for $N=5$ and $N=50$ vs.
$\mathcal{E}_{\rm CSIS}(1)$ and $C(5,1)$.}
\label{Fig:prac_N5_N50_p15_SNR01_P_T1_20}
\end{figure}

The schemes introduced in Section \ref{ssec:prac_scheme} for the
known CSIS case are simulated in Figure
\ref{Fig:prac_N5_N50_p15_SNR01_P_T1_20}. The performance of these
schemes are evaluated for $N = 5$ and $N = 50$. The performance of
these systems is compared against a lower bound given by
$\mathcal{E}_{\rm CSIS}(1)$ from (\ref{eqn:opt_err_exp_1_Rayleigh})
and an upper-bound, $C(5,K)$ from (\ref{eqn:CBM_choice}). The hybrid
scheme from Section \ref{sssec:hybrid_scheme} selects the better of
the two practical methods depending on the value of $\gamma_{\rm
s}$. It can be seen that even with these simple sub-optimal
practical schemes, the hybrid scheme is always better than
$\mathcal{E}_{\rm CSIS}(1)$, indicating that it is possible to
obtain multiple antenna gain. However, for each $N$, the hybrid
scheme does not approach the upper-bound of $C(5,K)$. When $N = 5$,
this is an expected result, since firstly, $C(N,K)$ is a bound that
is not necessarily achievable, and secondly, the practical schemes
are obtained as sub-optimal approximations to the optimal scheme
with full CSIS. The hybrid scheme for $N=50$ provides more gain over
$\mathcal{E}_{\rm CSIS}(1)$ than the hybrid scheme for $N=5$, but
does not beat $C(5,K)$. This means that although gains are possible
with the practical schemes, large gains are not possible, as
predicted by the bounds in Section \ref{ssec:full_CSIS}. For the
hybrid scheme, Method I is better at low values of $\gamma_{\rm s}$
and Method II is better at high values of $\gamma_{\rm s}$. The
value of $\gamma_{\rm s}$ at which the hybrid scheme changes methods
can also be seen in the simulations. In Figure
\ref{Fig:prac_N5_N50_p15_SNR01_P_T1_20}, the system has a channel
SNR, $\gamma_{\rm c} = 10$ dB, $p_{1} = 0.5$ and the Ricean-$K$
parameter is one. When there are five antennas at the FC, the hybrid
scheme changes from Method I to Method II at $\gamma_{\rm s} \approx
3$ dB, and when $N = 50$, the change occurs at $\gamma_{\rm s}
\approx 8.25$ dB. It can be seen that the hybrid scheme changes from
Method I to Method II at different values of $\gamma_{\rm s}$ based
on the system parameters. It can also be seen that when Method I is
selected by the hybrid scheme, the error in performance between
Method I and Method II is small. However, when Method II is selected
by the hybrid scheme, the performance gap between Method I and
Method II increases rapidly as $\gamma_{\rm s}$ increases.
Therefore, an underestimation of the value of $\gamma_{\rm s}$ is
tolerable, while an overestimation is not.

\begin{figure}[tb]
\begin{minipage}[b]{1.0\linewidth}
  \centering
  \centerline{\epsfig{figure=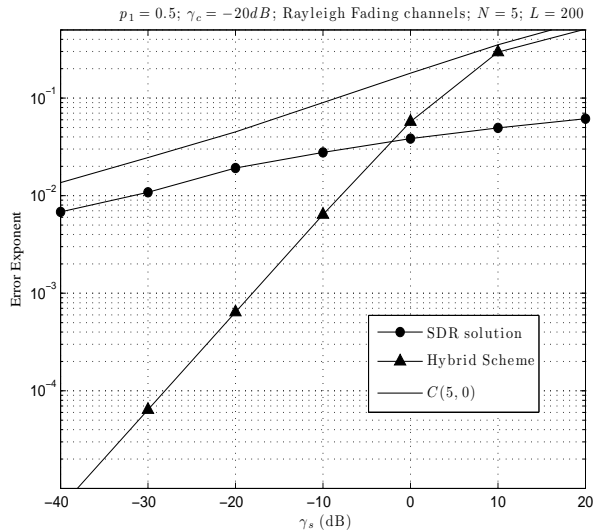,width=9cm,height=7.5cm}}
\end{minipage}
\caption{Hybrid realizable scheme, SDR relaxation and $C(N,K)$ vs $\gamma_{\rm s}$.}
\label{Fig:sdr_hybrid}
\end{figure}

The semidefinite relaxation (SDR) approach in Section \ref{sssec:sdr} is compared against the hybrid scheme (Section \ref{sssec:hybrid_scheme}) in Fig \ref{Fig:sdr_hybrid}. For the SDR solution, the value of $\mathbf{X}_{\rm OPT}$ from (\ref{eqn:po_sdr}) is calculated using \texttt{CVX}, a package for specifying and solving convex programs in MATLAB \cite{cvx}. It can be seen from these simulations that for low values of sensing SNR, $\gamma_{\rm s}$, the SDR solution outperforms the hybrid scheme. However, as the value of $\gamma_{\rm s}$ begins to increase, the hybrid scheme (which is designed to be optimal as $\gamma_{\rm s} \to\infty$) outperforms the SDR solution. The comparison with the upper-bound on the optimal error exponent, $C(N,K)$ is tight with respect to the better of the hybrid and SDR approaches. In order to obtain the solution to the SDR problem, an eigenvalue decomposition of $\mathbf{X}_{\rm OPT}$ is required, which is an $O(L^{3})$ operation \cite{Golub}. The SDR outperforms the hybrid scheme when $\gamma_{\rm s}$ is small, at the expense of increased complexity. 

\section{Conclusions}
\label{sec:conclusions}

A distributed detection system with sensors transmitting
observations to a fusion center with multiple antennas is
considered. The error exponent is derived from the conditional
probability of error. AWGN and fading channels are considered
between the sensors and the fusion center with varying amounts of
CSI at the sensors. A large number of sensors and a finite number of
antennas, or a large number of sensors and antennas are considered.
The gain due to having multiple antennas, defined in terms of the
ratio of error exponents relative to the single antenna case for
large $L$ is quantified. The asymptotic nature of the sensors and
the corruption of the sensed data with noise are the two major
distinguishing factors in this work, compared to the conventional
analysis of point-to-point MIMO systems.

The gains due to the number of antennas at the FC, both when CSI is
available and unavailable at the sensor side, are now summarized. It
is seen that when $K>0$, when CSIS is unavailable, and when
$\gamma_{\rm c} = 0$, the gain in error exponent can grow unbounded
with $N$. When the channel is zero-mean (such as the Rayleigh case
with $K \to 0$), and when CSIS is not available, adding antennas
improves the error exponent linearly. However, when CSIS is
available, the gain is bounded by $2 (E[|h_{l}|])^{-2}$. In the
special case of Rayleigh fading channels, when CSIS is available,
the gain due to multiple antennas at the FC is bounded by $8/\pi$.
As both $N$ and $L$ increase, the error exponents for $K>0$ are
infinite, yielding an undefined gain. For zero-mean channels the
dependence on $\beta$ from (\ref{eqn:LNratio}) indicates an
increasing gain when $\beta$ is small ($L \ll N$), and bounded gain
when $\beta$ is large ($L \gg N$). When the sensors use only channel
phase information, the gain due to multiple antennas at the FC is
upper bounded by $(E[|h_{l}|])^{-2}$, irrespective of the number of
antennas at the FC.

In the case when full CSIS is available, it has been argued that
having multiple antennas at the FC provides limited gains, which can
be exploited by sub-optimal schemes. In one approach, the system is
configured to beamform to the antenna that provides the best
performance, where the FC still uses the data gathered at the other
antennas. On an average, this is shown to perform better than in the
single antenna case. Another approach is to assume there is no
sensing noise, and set sensor gains tuned for such a system even
when sensing noise is present. In this case, the system performs
optimally when the sensing noise in the system is low. A hybrid
scheme is proposed which selects the better of these two methods
depending on the sensing SNR, $\gamma_{\rm s}$.

Depending on the number of sensors and antennas at the FC, and their
rates of growth, the following system design recommendation can be
made. If CSIS is available and $N\ll L$, then for better
performance, it is recommended to increase the number of sensors,
rather than the number of antennas at the FC. However, if $N$ can be
increased at a much faster rate than $L$, it is possible to achieve
greater gains due to adding antennas at the FC.

\appendices

\section{Proof of Theorem \ref{thm:corr_obs_noise}}
\label{app:proof_corr_obs_noise}

We begin by noting that the presence of correlation in $\eta_{l}$
affects the total average transmit power. Therefore, to prove
Theorem \ref{thm:corr_obs_noise}, we need to reconsider the
following in presence of correlation: (i) the power constraint; (ii)
the AWGN upper-bound in (\ref{eqn:A_bound}); (iii) the ``no sensing
noise'' upper-bound in (\ref{eqn:B_bound}), which will then be used
to redefine the combined upper-bound in (\ref{eqn:CBM_choice}).

(i) Power constraint: The total transmitted power is given by
\begin{equation}
P_{T} = E \left[ \sum_{l=1}^{L} \left| \alpha_{l} \left( \Theta +
\eta_{l} \right) \right|^{2} \right] = \boldsymbol{\alpha}^{H}
\left( p_{1} \theta^{2} \mathbf{I}_{L} + \mathbf{R}_{\eta} \right)
\boldsymbol{\alpha}, \label{eqn:tx_power_corr}
\end{equation}
and constrained as
\begin{equation}
\boldsymbol{\alpha}^{H} \left( p_{1} \theta^{2} \mathbf{I}_{L} +
\mathbf{R}_{\eta} \right) \boldsymbol{\alpha} \leq P_{T}.
\label{eqn:tx_power_corr_ineq}
\end{equation}
If (\ref{eqn:tx_power_corr_ineq}) holds, then
\begin{equation}
\| \boldsymbol{\alpha} \|^{2} \leq \frac{P_{T}} {p_{1} \theta^{2} +
\lambda_{\rm min}} \mathop{:=} P, \label{eqn:P_constr}
\end{equation}
also holds. Since (\ref{eqn:P_constr}) is less stringent than
(\ref{eqn:tx_power_corr_ineq}), if (\ref{eqn:P_constr}) is used
instead of the original power constraint in
(\ref{eqn:tx_power_corr_ineq}), an upper-bound will be obtained in
the subsequent derivation of the error exponent.

(ii) Upper-bound (AWGN channels): Recall that in this case,
$\mathbf{H} = \boldsymbol{1}_{N \times L}$. Since the sensing noise
is not iid, $\boldsymbol{\alpha}$ has to be selected in such a way
that the error exponent is maximized:
\begin{equation}
\maximize_{\boldsymbol{\alpha}} \quad \boldsymbol{\alpha}^{H}
\boldsymbol{1}_{L\times N} \mathbf{R} (\boldsymbol{\alpha})^{-1}
\boldsymbol{1}_{N\times L}
\boldsymbol{\alpha} \quad \text{subject to } \quad  \boldsymbol{\alpha}^{H}
\boldsymbol{\alpha} \leq P, \label{eqn:thm_awgn_opt_alpha}
\end{equation}
to yield the error exponent in the AWGN case with correlated sensing
noise:
\begin{equation}
\mathcal{E}_{\rm AWGN} (N) \leq \frac{1} {L} \frac{\theta^{2}}
{8}(\boldsymbol{\alpha}_{\rm AWGN}^{\rm opt})^{H}
\boldsymbol{1}_{L\times N} \mathbf{R} (\boldsymbol{\alpha}_{\rm
AWGN}^{\rm opt})^{-1} \boldsymbol{1}_{N\times L}
\boldsymbol{\alpha}_{\rm AWGN}^{\rm opt},
\label{eqn:thm_awgn_err_exp_setup}
\end{equation}
where $\boldsymbol{\alpha}_{\rm AWGN}^{\rm opt}$ provides to
solution to (\ref{eqn:thm_awgn_opt_alpha}) and the inequality in
(\ref{eqn:thm_awgn_err_exp_setup}) is due to (\ref{eqn:err_exp_fn})
and the modified power constraint in (\ref{eqn:P_constr}).
To fully compute an upper bound on the right hand side of (\ref{eqn:thm_awgn_err_exp_setup}), first, $\mathbf{R}(\boldsymbol{\alpha})$ is inverted
and simplified. For the case of correlated noise,
$\mathbf{R}(\boldsymbol{\alpha})$ is given by
\begin{align}
\mathbf{R}(\boldsymbol{\alpha}) &= \boldsymbol{1}_{N\times L}
\mathbf{D} (\boldsymbol{\alpha}) \mathbf{R}_{\eta} \mathbf{D}
(\boldsymbol{\alpha})^{H} \boldsymbol{1}_{L\times N} +
\sigma_{\nu}^{2}
\mathbf{I}_{N} \nonumber \\
& \succeq \lambda_{\rm min} \boldsymbol{1}_{N \times L} \mathbf{D}
(\boldsymbol{\alpha})^{H} \mathbf{D} (\boldsymbol{\alpha})
\boldsymbol{1}_{L \times N} + \sigma_{\nu}^{2} \mathbf{I}_{N},
\label{eqn:thm_corr_R}
\end{align}
where $\mathbf{A} \succeq \mathbf{B}$ indicates that the matrix
($\mathbf{A} - \mathbf{B}$) is positive semi-definite. Using the
Sherman-Morrison-Woodbury formula for matrix inversion,
\begin{equation}
\mathbf{R} (\boldsymbol{\alpha})^{-1} \preceq \frac{1}
{\sigma_{\nu}^{2}} \mathbf{I}_{N} - \frac{1} {\sigma_{\nu}^{2}}
\boldsymbol{1}_{N \times L} \left[ \text{diag} \left( \frac{1}
{\lambda_{\rm min} |\alpha_{i}|^{2}} \right) +
\frac{\boldsymbol{1}_{L \times N} \boldsymbol{1}_{N \times L}}
{\sigma_{\nu}^{2}} \right]^{-1} \boldsymbol{1}_{L \times N} \frac{1}
{\sigma_{\nu}^{2}}. \label{eqn:thm_r_inv_1}
\end{equation}
Invoking the Sherman-Morrison-Woodbury formula for matrix inversion
once again,
\begin{equation}
\mathbf{R} (\boldsymbol{\alpha})^{-1} \preceq \frac{1}
{\sigma_{\nu}^{2}} \mathbf{I}_{N} - \frac{1}{\sigma_{\nu}^{2}} M
\boldsymbol{1}_{N\times N}, \label{eqn:r_inv}
\end{equation}
where
\begin{equation}
M \mathop{:=} \frac{\displaystyle{\sum_{l=1}^{L} \frac{\lambda_{\rm min}}
{\sigma_{\nu}^{2}} |\alpha_{l}|^{2}}} {\displaystyle{1 + N
\sum_{i=1}^{L}} \frac{\lambda_{\rm min}} {\sigma_{\nu}^{2}}
|\alpha_{i}|^{2}} \leq \frac{\lambda_{\rm min} P} {N \lambda_{\rm
min} P + \sigma_{\nu}^{2}}, \label{eqn:thm_M}
\end{equation}
due to the fact that $\boldsymbol{\alpha}^{H} \boldsymbol{\alpha}
\leq P$ from (\ref{eqn:P_constr}).

By substituting (\ref{eqn:r_inv}) in (\ref{eqn:thm_awgn_opt_alpha}),
the solution to (\ref{eqn:thm_awgn_opt_alpha}) is upper-bounded by
the solution to
\begin{equation}
\maximize_{\boldsymbol{\alpha}} \quad \boldsymbol{\alpha}^{H}
\boldsymbol{1}_{L\times L} \boldsymbol{\alpha} \quad \text{ subject to } \quad  \boldsymbol{\alpha}^{H}
\boldsymbol{\alpha} \leq P. \label{eqn:thm_awgn_prbm_red}
\end{equation}
The value of $\boldsymbol{\alpha}$ that maximizes
(\ref{eqn:thm_awgn_prbm_red}) is the eigenvector corresponding to
the maximum eigenvalue of $\boldsymbol{1}_{L\times L}$, scaled to
satisfy the constraint with equality. Substituting this in
(\ref{eqn:thm_awgn_err_exp_setup}), the bound in (\ref{eqn:A_bound})
obtained, with the substitution, $\gamma_{\rm s} =
\tilde{\gamma}_{\rm s}$, where $\tilde{\gamma}_{\rm s} = \theta^{2}/
\lambda_{\rm min}$ and $P_{T} \leq P/(p_{1} \theta^{2} +
\lambda_{\rm min})$.

(iii) Upper-bound (no sensing noise): With no sensing noise,
$\mathbf{R}_{\eta} = \boldsymbol{0}_{L\times L}$. The optimization
problem to obtain the best error exponent is the same as in
(\ref{eqn:full_CSIS_no_obs_noise_setup}), to yield
(\ref{eqn:B_bound}).

Combining the modified AWGN upper-bound and the no sensing noise
upper-bound in (\ref{eqn:B_bound}), a joint upper-bound is obtained,
which is identical to (\ref{eqn:CBM_choice}), except for the
substitution $\sigma_{\eta}^{2} = \lambda_{\rm min}$ and
$\tilde{\gamma}_{\rm s} = \theta^{2}/\lambda_{\rm min}$. It follows
that (\ref{eqn:thm_gain_opt}) holds with $z = \gamma_{\rm c}/ (p_{1}
\tilde{\gamma}_{\rm s} + 1)$, to provide the proof.

\bibliography{multiant_bib}

\begin{thebibliography}{10}
\providecommand{\url}[1]{#1}
\csname url@samestyle\endcsname
\providecommand{\newblock}{\relax}
\providecommand{\bibinfo}[2]{#2}
\providecommand{\BIBentrySTDinterwordspacing}{\spaceskip=0pt\relax}
\providecommand{\BIBentryALTinterwordstretchfactor}{4}
\providecommand{\BIBentryALTinterwordspacing}{\spaceskip=\fontdimen2\font plus
\BIBentryALTinterwordstretchfactor\fontdimen3\font minus
  \fontdimen4\font\relax}
\providecommand{\BIBforeignlanguage}[2]{{%
\expandafter\ifx\csname l@#1\endcsname\relax
\typeout{** WARNING: IEEEtran.bst: No hyphenation pattern has been}%
\typeout{** loaded for the language `#1'. Using the pattern for}%
\typeout{** the default language instead.}%
\else
\language=\csname l@#1\endcsname
\fi
#2}}
\providecommand{\BIBdecl}{\relax}
\BIBdecl

\bibitem{viswanathan}
R.~Viswanathan and P.~Varshney, ``Distributed detection with multiple sensors
  {I. Fundamentals},'' \emph{Proceedings of the IEEE}, vol.~85, no.~1, pp.
  54--63, January 1997.

\bibitem{xiao05}
J.-J. Xiao and Z.-Q. Luo, ``Universal decentralized detection in a
  bandwidth-constrained sensor network,'' \emph{IEEE Transactions on Signal
  Processing}, vol.~53, no.~8, pp. 2617 -- 2624, August 2005.

\bibitem{banavar09}
M.~K. Banavar, C.~Tepedelenlio\u{g}lu, and A.~Spanias, ``Estimation over fading
  channels with limited feedback using distributed sensing,'' \emph{IEEE
  Transactions on Signal Processing}, vol.~58, no.~1, pp. 414--425, January
  2010.

\bibitem{Chamberland2004}
J.-F. Chamberland and V.~V. Veeravalli, ``Asymptotic results for decentralized
  detection in power constrained wireless sensor networks,'' \emph{IEEE Journal
  on Selected Areas in Communications}, vol.~22, no.~6, pp. 1007--1015, August
  2004.

\bibitem{Chen04}
B.~Chen, R.~Jiang, T.~Kasetkasem, and P.~K. Varshney, ``Channel aware decision
  fusion in wireless sensor networks,'' \emph{IEEE Transactions in Signal
  Processing}, vol.~52, no.~12, pp. 3454--3458, December 2004.

\bibitem{Li07}
W.~Li and H.~Dai, ``Distributed detection in wireless sensor networks using a
  multiple access channel,'' \emph{IEEE Transactions on Signal Processing},
  vol.~55, no.~3, pp. 822--833, March 2007.

\bibitem{Wimalajeewa08}
T.~Wimalajeewa and S.~K. Jayaweera, ``Optimal power scheduling for correlated
  data fusion in wireless sensor networks via constrained {PSO},'' \emph{IEEE
  Transactions on Wireless Communications}, vol.~7, no.~9, pp. 3608--3618,
  September 2008.

\bibitem{Evans08a}
F.~Li and J.~S. Evans, ``Design of distributed detection schemes for
  multiaccess channels,'' \emph{Communications Theory Workshop, 2008. AusCTW
  2008. Australian}, pp. 51--57, January 2008.

\bibitem{Chen96}
J.~Chen, N.~Ansari, and Z.~Siveski, ``Distributed detection for cellular
  {CDMA},'' \emph{Electronics Letters}, vol.~32, no.~3, pp. 169--171, February
  1996.

\bibitem{Zhang08}
X.~Zhang, H.~V. Poor, and M.~Chiang, ``Optimal power allocation for distributed
  detection over {MIMO} channels in wireless sensor networks,'' \emph{IEEE
  Transactions on Signal Processing}, vol.~56, no.~9, pp. 4124--4140, September
  2008.

\bibitem{Liu2007}
K.~Liu and A.~M. Sayeed, ``Type-based decentralized detection in wireless
  sensor networks,'' \emph{IEEE Transactions on Signal Processing}, vol.~55,
  no.~5, pp. 1899--1910, May 2007.

\bibitem{Anandkumar2007}
A.~Anandkumar and L.~Tong, ``Type-based random access for distributed detection
  over multiaccess fading channels,'' \emph{IEEE Transactions on Signal
  Processing}, vol.~55, no.~10, pp. 5032--5043, October 2007.

\bibitem{Berger2009}
C.~R. Berger, M.~Guerriero, S.~Zhou, and P.~Willett, ``{PAC} vs. {MAC} for
  decentralized detection using noncoherent modulation,'' \emph{IEEE
  Transactions in Signal Processing}, vol.~57, no.~9, pp. 3562--3575, September
  2009.

\bibitem{Yiu2009}
S.~Yiu and R.~Schober, ``Nonorthogonal transmission and noncoherent fusion of
  censored decisions,'' \emph{IEEE Transactions on Vehicular Technology},
  vol.~58, no.~1, pp. 263--273, January 2009.

\bibitem{Jayaweera05}
S.~K. Jayaweera, ``Large system decentralized detection performance under
  communication constraints,'' \emph{IEEE Communications Letters}, vol.~9,
  no.~9, pp. 769--771, September 2005.

\bibitem{Jayaweera07}
------, ``Bayesian fusion performance and system optimization for distributed
  stochastic {Gaussian} signal detection under communication constraints,''
  \emph{IEEE Transactions on Signal Processing}, vol.~55, no.~4, pp.
  1238--1250, April 2007.

\bibitem{Tarzai05}
K.~A.~A. Tarzai, S.~K. Jayaweera, and V.~Aravinthan, ``Performance of
  decentralized detection in a resource-constrained sensor network with
  non-orthogonal communications,'' \emph{In Proc. 39th Annual Asilomar
  Conference on Signals, Systems and Computers}, pp. 437--441, October 2005.

\bibitem{Bai08}
K.~Bai and C.~Tepedelenlio\u{g}lu, ``Distributed detection in {UWB} wireless
  sensor networks,'' \emph{Proceedings of the IEEE ICASSP 2008}, pp.
  2261--2264, April 2008.

\bibitem{Golub}
G.~H. Golub and C.~F.~V. Loan, \emph{Matrix computations}, 3rd~ed.\hskip 1em
  plus 0.5em minus 0.4em\relax Baltimore, MD: John Hopkins University Press,
  1996.

\bibitem{Merg06}
G.~Mergen and L.~Tong, ``Type based estimation over multiaccess channels,''
  \emph{IEEE Transactions on Signal Processing}, vol.~54, no.~2, pp. 613--626,
  February 2006.

\bibitem{Liu2007a}
K.~Liu, H.~E. Gamal, and A.~Sayeed, ``Decentralized inference over
  multiple-access channels,'' \emph{IEEE Transactions on Signal Processing},
  vol.~55, no.~7, pp. 3445--3455, July 2007.

\bibitem{xiao08}
J.~Xiao, S.~Cui, Z.-Q. Luo, and A.~J. Goldsmith, ``{Linear coherent
  decentralized estimation},'' \emph{IEEE Transactions on Signal Processing},
  vol.~56, no.~2, pp. 757--770, February 2008.

\bibitem{Boyd}
S.~Boyd and L.~Vandenberghe, \emph{Convex Optimization}.\hskip 1em plus 0.5em
  minus 0.4em\relax New York: Cambridge University Press, 2004.

\bibitem{Integrals2007}
I.~S. Gradshteyn and I.~M. Ryzhik, \emph{Table of Integrals, Series and
  Products}, 7th~ed.\hskip 1em plus 0.5em minus 0.4em\relax UK: Elsevier, 2007.

\bibitem{Abra65}
M.~Abramowitz and I.~A. Stegun, \emph{Handbook of Mathematical
  Functions}.\hskip 1em plus 0.5em minus 0.4em\relax Courier Dover
  Publications, 1965.

\bibitem{Porat93}
B.~Porat, \emph{Digital processing of random signals: theory and
  methods}.\hskip 1em plus 0.5em minus 0.4em\relax New Jersey: Prentice-Hall,
  1993.

\bibitem{Sendonaris03a}
A.~Sendonaris, E.~Erkip, and B.~Aazhang, ``User cooperation diversity - {Part
  I: System} description,'' \emph{IEEE Transactions on Communications},
  vol.~51, pp. 1927--1938, November 2003.

\bibitem{Verdu04}
A.~M. Tulino and S.~Verdu, \emph{Random Matrix Theory and Wireless
  Communications}.\hskip 1em plus 0.5em minus 0.4em\relax USA: {now
  Publishers}, 2004.

\bibitem{Yin88}
Y.~Q. Yin, Z.~D. Bai, and P.~R. Krishnaiah, ``On the limit of the largest
  eigenvalue of the large dimensional sample covariance matrix,''
  \emph{Probability Theory and Related Fields}, vol.~78, no.~4, pp. 509--521,
  August 1988.

\bibitem{Bai93}
Z.~D. Bai and Y.~Q. Yin, ``Limit of the smallest eigenvalue of a large
  dimensional sample covariance matrix,'' \emph{The Annals of Probability},
  vol.~21, no.~3, pp. 1275--1294, 1993.

\bibitem{Luo2010}
Z.-Q. Luo, W.-K. Ma, A.~M.-C. So, Y.~Ye, and S.~Zhang, ``Semidefinite
  relaxation of quadratic optimization problems,'' \emph{IEEE Signal Processing
  Magazine}, vol.~27, no.~3, pp. 20--34, May 2010.

\bibitem{cvx}
M.~Grant and S.~Boyd, ``{CVX}: Matlab software for disciplined convex
  programming, version 1.21,'' \url{http://cvxr.com/cvx}, Jul. 2010.

\end{thebibliography}
\end{document}